\documentclass{aa}  

\usepackage{color}

\usepackage{graphicx}
\usepackage{txfonts}

\usepackage{amssymb}
\usepackage{multirow}
\usepackage{pifont}
\usepackage{diagbox}

\makeatletter
    
    \newcommand{\Rmnum}[1]{\expandafter\@slowromancap\romannumeral #1@}
\makeatother
\begin{document}

      
      \title{The invasion of a free floating planet and the number asymmetry of Jupiter Trojans}


     \author{Jian Li\inst{1,2}
           \and
          Zhihong Jeff Xia\inst{3} 
          \and
          Nikolaos Georgakarakos\inst{4,5}
          \and
          Fumi Yoshida\inst{6,7}
          }

   \institute{School of Astronomy and Space Science, Nanjing University, 163 Xianlin Avenue, Nanjing 210023, PR China\\
              \email{ljian@nju.edu.cn}
         \and
         Key Laboratory of Modern Astronomy and Astrophysics in Ministry of Education, Nanjing University, Nanjing 210023, PR China
        \and
             Department of Mathematics, Northwestern University, 2033 Sheridan Road, Evanston, IL  60208, USA
         \and New York University Abu Dhabi, PO Box 129188 Abu Dhabi, United Arab Emirates
         \and Center for Astro, Particle and Planetary Physics (CAP3), New York University Abu Dhabi, PO Box 129188 Abu Dhabi, United Arab Emirates
         \and
         University of Occupational and Environmental Health, 1-1 Iseigaoka, Yahata, Kitakyusyu 807-8555, Japan
         \and
         Planetary Exploration Research Center, Chiba Institute of Technology, 2-17-1 Tsudanuma, Narashino, Chiba 275-0016, Japan \\
             }
   \date{Received September 15, 1996; accepted March 16, 1997}

 
  \abstract
   {This paper extends our previous study \citep[][]{li2023} of the early evolution of Jupiter and its two Trojan swarms by introducing the possible perturbations of a free floating planet (FFP) invading the Solar System.}
   {In the framework of the invasion of a FFP, we aim to provide some new scenarios to explain the number asymmetry of the L4 and L5 Jupiter Trojans, as well as some other observed features (e.g. the resonant amplitude distribution).}
    {We investigate two different cases: (i) The indirect case, where Jupiter experiences a scattering encounter with the FFP and jumps outwards at a speed that is much higher than that considered in \citet[][]{li2023}, resulting in a change in the numbers of the L4 ($N_4$) and L5 ($N_5$) Trojans swarms. (ii) The direct case, in which the FFP traverses the L5 region and affects the stability of the local Trojans.}
   {In the indirect case, the outward migration of Jupiter can be fast enough to make the L4 islands disappear temporarily, inducing a resonant amplitude increase of the local Trojans. After the migration is over, the L4 Trojans come back to the re-appeared and enlarged islands. As for the L5 islands, they always exist but expand even more considerably. Since the L4 swarm suffers less excitation in the resonant amplitude than the L5 swarm, more L4 Trojans are stable and could survive to the end. In the direct case, the FFP could deplete a considerable fraction of the L5 Trojans, while the L4 Trojans at large distances are not affected and all of them could survive.} 
   {Both the indirect and direct cases could result in a number ratio of $R_{45}=N_4/N_5\sim1.6$ that can potentially explain the current observations. The latter has the advantage of producing the observed resonant amplitude distribution. For achieving these results, we propose that the FFP should have a mass of at least of a few tens of Earth masses and its orbital inclination is allowed to be as high as $40^{\circ}$.}
   {}

   
   \titlerunning{Asymmetry in the number of L4 and L5 Jupiter Trojans}

   \maketitle
%

\section{Introduction}

In 1772, J. L. Lagrange showed that there are two equilibrium points in the orbital plane of two massive bodies where small mass bodies can remain around.  Later in 1906, Max Wolf at the Heidelberg observatory discovered an asteroid (588 Achilles) at a distance from the Sun similar to that of Jupiter \citep{nich61}. Since then, more and more asteroids that share the orbit of Jupiter were discovered. As Lagrange's theory predicted, there are two swarms of these asteroids, leading or trailing Jupiter by about $60^{\circ}$ in longitude, that is, around the L4 or L5 triangular Lagrangian point. The asteroids ahead of Jupiter are called the L4 Jupiter Trojans, while the others behind Jupiter are the L5 Jupiter Trojans.  

For decades, people have known that are many more objects in the L4 swarm than in the L5 one \citep{shoe89, jewi04, frei06, naka08, grav11, grav12, slyu13, li2018}. This number discrepancy still holds today, even when more than 11000 Jupiter Trojans have been discovered. The unbiased leading-to-trailing number ratio is estimated to be about 1.6 \citep{szab07}. In the current configuration of the Solar System, the L4 and L5 Trojan swarms show comparable dynamical stability and survivability properties \citep{disi14, disi19, holt20}. Thus, the significant number difference should arise at earlier times of our Solar System's life, and it remains a profound mystery in the evolutionary history of Jupiter Trojans.

In a very recent work \citep{li2023}, we have developed a new mechanism that may explain the number asymmetry of Jupiter Trojans. We proposed that an outward migration of Jupiter, caused by the giant planet instability in the early Solar System, can induce different evolutional paths for the two Trojan swarms. In this dynamical model, the L4 Trojans can have their resonant amplitudes decreased due to the contraction of the L4 region and they would become more stable. Oppositely, the L5 Trojans' resonant amplitudes increase because of the expansion of the L5 region and therefore they would become more unstable. As a result, there are more surviving Trojans in the L4 swarm than in the L5 one. As long as the outward migration of Jupiter is fast enough, at a speed of $\dot{a}_J\sim1.5\times10^{-4}$ AU/yr ($\dot{a}_J$ denotes the variation of Jupiter's semimajor axis $a_J$), this mechanism could provide a natural explanation for the unbiased observation, that the L4 Trojans are about 1.6 times more than the asteroids in the L5 swarm. 

The fast outward migration of Jupiter can always be capable of inducing a leading-to-trailing number ratio $R_{45}>1$, and theoretically speaking, the resulting L4/L5 asymmetry should be even more significant as long as the migration of Jupiter is faster. Regarding this rapid migration of Jupiter, in \citet{li2023}, we considered the mechanism of early dynamical instability among the giant planets in the outer Solar System (see \citet{nesy08} for a review). When Jupiter experienced a close encounter with the fifth outer planet, which left the Solar System eventually, it could migrate outwards at a speed about $1.5\times10^{-4}$ AU/yr \citep{nesy13}, as we mentioned above. For this dynamical model confined within the Solar System, the migration speed of Jupiter is limited. Thus, in order to achieve the number asymmetry of $R_{45}\sim1.6$, we have to assume that the Trojans initially possess moderately large resonant amplitudes of $\ge30^{\circ}$.


In this paper, we extend the previous work by considering a much faster migration of Jupiter, which could be produced by the invasion of a free floating planet (FFP). Observations have shown that the number of FFPs with Jupiter-mass can exceed that of Main-Sequence stars in our Galaxy \citep{sumi11}. Although the probability of capturing a FFP is only a few per cent or even lower in planetary systems \citep{lig16, park17, goul18}, a penetrating flyby in the Solar System could be more likely. If a FFP happened to interact with Jupiter, depending on the minimum mutual distance, Jupiter could migrate outwards by a considerable distance. Since the timescale of this process is very short and nearly fixed for various FFPs with perihelia around the orbit of Jupiter, as we will show in Sect. 2, the migration speed of Jupiter can be quite high.

According to \citet{sica03}, there is a critical migration speed of $\dot{a}_J^{crit}=3.8\times10^{-3}$ AU/yr. The jumping-Jupiter models constructed in \citet{li2023} exclusively consider the speed values of $\dot{a}_J$ below this limit, indicating the persistent existence of the L4 islands and thus the possible survival of the local Trojans. But, due to the gravitational acceleration of the FFP, Jupiter could jump outwards at a speed of $\dot{a}_J>\dot{a}_J^{crit}$, leading to the disappearance of the L4 islands. Thus, it is of great interest to understand how the L4 Trojans would be influenced. 

Besides the fast outward migration of Jupiter caused by the FFP, a second potential mechanism of producing the number asymmetry of Jupiter Trojans in the case that a FFP once invaded the L5 region is also explored. Actually, \citet{nesy13} proposed that after Jupiter Trojans are captured, an extra ice giant can still remain on the Jupiter-crossing orbit. If this ice giant occasionally traverses the L5 region, a fraction of the local Trojans would be scattered away. Since the other group of Trojans in the L4 region are far away from this ice giant, their stability is not affected. Obviously, a L4/L5 number ratio of $R_{45}>1$ would appear. The authors found that in one case, $R_{45}$ can reach a value of about 1.3, which seems not large enough to explain the current asymmetry of $R_{45}\sim1.6$. Possibly, the number asymmetry generated in this scenario could be more significant if the perturbations on the L5 Trojans are stronger, for instance, by increasing the mass of the invading planet. However, in the model of \citet{nesy13}, the planet invading the L5 region cannot have sufficiently large mass because it is an ice giant and should have a mass comparable to that of Uranus/Neptune. But if we consider the invading planet to be a FFP, it could be much more massive, up to the order of Jupiter mass.

In this paper, as a follow-up work of our study of the number asymmetry of Jupiter Trojans, we investigate the influence of a FFP invading the Solar system. The rest of this paper is organised as follows: in Sect. 2, we investigate the outward migration of Jupiter at a very high speed due to the close encounter with the FFP, and the concomitant evolution of Jupiter Trojans that may explain the unbiased L4/L5 number ratio of $R_{45}\sim1.6$. In Sect. 3, we investigate another potential mechanism of producing the number asymmetry of Jupiter Trojans in the case that the FFP once penetrated the L5 region and dispersed a large fraction of the local population. In Sect. 4, we summarise the results and provide some discussions.

\section{Indirect case: Jupiter jumping caused by a FFP}

As we described in the introduction, a FFP could visit the Solar System on an unbound orbit and possibly pass close to Jupiter. During the time of such close encounter, 
the FFP could accelerate Jupiter's motion, inducing the increase of Jupiter's semimajor-axis by an amplitude of $\Delta a_J$. Due to the short period of the FFP's strong perturbation, the migration timescale $\Delta t$ of Jupiter should equivalently be very small. Therefore, the migration rate of Jupiter, that is $\dot{a}_J=\Delta{a}_J/\Delta{t}$, could be much higher than that considered in \citet{li2023}. Obviously, the reasonable values of $\Delta a_J$ and $\Delta t$ are to be determined by the orbit of the FFP.

\subsection{Orbit of the FFP}

\begin{figure}
 \hspace{0 cm}
  \includegraphics[width=8.5cm]{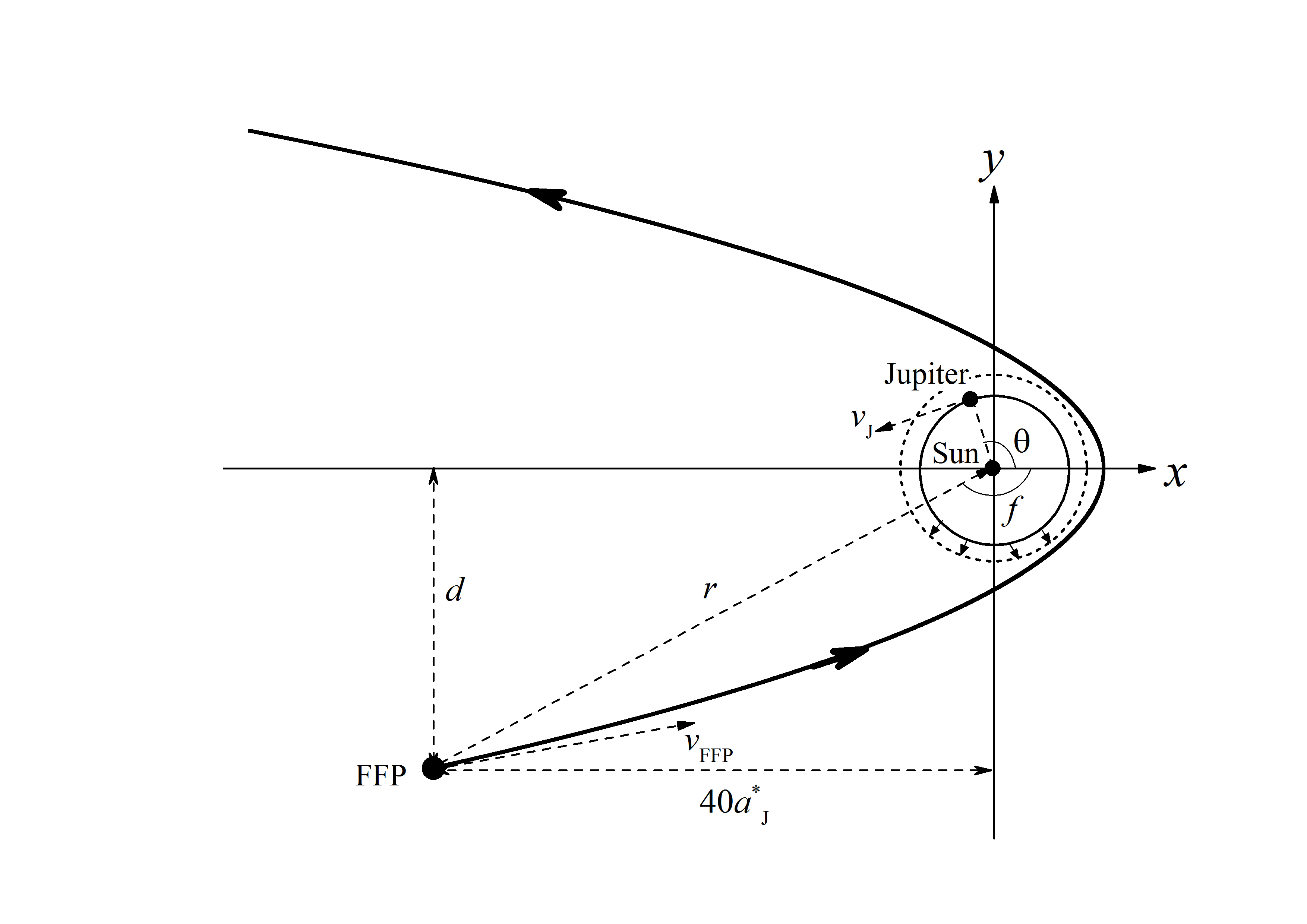}
  \caption{Schematic diagram of the jumping-Jupiter evolution due to the invasion of a FFP in the co-planar model. Jupiter is initially on a circular orbit (solid circle) around the Sun, with phase angle $\theta$ measured anticlockwise from the positive $x$-axis. The FFP is approaching the Sun-Jupiter system on a parabolic orbit with initial coordinates $(-40a_J^{\ast}, -d)$ ($d>0$). After the close encounter near the FFP's perihelion on the positive $x$-axis, Jupiter has its motion accelerated and its orbit expanded accordingly (dashed circle), while the FFP leaves the Solar System.}
  \label{3body}
\end{figure}

In order to investigate the FFP's gravitational effect on Jupiter's migration, we simply use a three-body system consisting of the Sun with mass $M_{\odot}$, Jupiter with mass $m_J$ and an incoming FFP with mass $m_{FFP}$ (see Fig. \ref{3body}). The Sun is fixed at the origin of the coordinate system $(x, y)$. In this heliocentric framework, Jupiter initially moves on a Keplerian circular orbit; and similar to \citet{varv12}, the FFP is introduced to start on a parabolic trajectory. We assume that the orbits of these three bodies are all in the same plane.

If we consider Jupiter to be moving anticlockwise around the Sun on a circular orbit, then, the initial positions and velocities of Jupiter can be expressed as
\begin{eqnarray}
x_J(0)&=&a_J^{\ast}\cos\theta, ~~~~~~~~~~ y_J(0)=a_J^{\ast}\sin\theta,\nonumber\\
\dot{x_J}(0)&=&-v_J(0)\sin\theta,~~~~~\dot{y}(0)=v_J(0)\cos\theta,
\label{initialJ}
\end{eqnarray}
where $a_J^{\ast}$ denotes the semimajor axis of the unperturbed orbit of Jupiter and is taken to be 5.2 AU, $\theta$ is Jupiter's initial position angle (anticlockwise from the positive $x$-axis), and the initial linear velocity can be calculated by
\begin{equation}
v_J(0)=\sqrt{\frac{G(M_{\odot}+m_J)}{a_J^{\ast}}}.
\end{equation}
Here, and throughout the following, the bracketed digit 0 indicates the initial value of each parameter at the time $t=0$.

The FFP is initially started at $(-40a_J^{\ast}, -d)$ ($d>0$), having a parabolic velocity of $v_{FFP}$ with respect to the Sun. Then the parameter $d$ controls the FFP's perihelion, which will be kept outside the orbit of Jupiter to avoid the collision event as we illustrate below. The orbit of the FFP can be described by 
\begin{equation}
r=\frac{p}{1+\cos f},
\label{radial}
\end{equation}
where $r$ is the radial distance, $f$ is the true anomaly, and $p$ is the semilatus rectum that can be determined by the initial coordinates. Then the initial conditions of the FFP can be written as
\begin{eqnarray}
x(0)&=&-40a_J^{\ast}, ~~~~~~~~~~~~~~~~~~ y(0)=-d,\nonumber\\
\dot{x}(0)&=&-\sqrt{{\tilde{\mu}}/{p}}\sin f(0),~~~\dot{y}(0)=\sqrt{{\tilde{\mu}}/{p}}~(1+\cos f(0)),
\label{initialFFP}
\end{eqnarray}
where $\tilde{\mu}=G(M_{\odot}+m_{FFP})$.

For the system's initial conditions described above by Eqs. (\ref{initialJ}) and (\ref{initialFFP}), there are three free parameters: $\theta$ for Jupiter, $d$ and $m_{FFP}$ for the FFP.  The initial phase angle $\theta$ determines the location of Jupiter when the FFP is close to Jupiter's orbit. The initial $y$-distance $d$ determines where exactly the FFP's perihelion ($p/2$, 0) is, and accordingly the minimum distance between the FFP and Jupiter. Finally, the mass $m_{FFP}$ determines the strength of the FFP's perturbation from a certain distance. Therefore, the gravitational effect of the FFP on Jupiter could be very sensitive to these three parameters.

First, in order to avoid a collision between Jupiter and the FFP, we need to adopt the value of $d$ satisfying the condition $p/2>a_J^{\ast}$. Let $r(0)$ be the initial radial distance of the FFP, we have
\begin{equation}
p=r(0)\cdot(1+\cos f(0))=r(0)+x(0)>2a_J^{\ast},
\label{pvalue}
\end{equation}
where
\begin{equation}
r(0)=\sqrt{(x(0))^2+(y(0))^2}.
\label{r0value}
\end{equation}
Substituting for $x(0)$ and $y(0)$ from Eq. (\ref{initialFFP}) into Eqs. (\ref{pvalue}) and (\ref{r0value}), we get
\begin{equation}
d>12.8a_J^{\ast}.
\label{dvalue}
\end{equation}
This condition can guarantee that the FFP will always be outside the orbit of Jupiter and supply the perturbing acceleration for Jupiter's migration via a close encounter, as visualised in Fig. \ref{3body}.

Second, in order for a close encounter to be achieved, it is required that the FFP comes close to the actual position of Jupiter and not only to its orbit. Having this in mind, we would design the configuration in a way that both of the FFP and Jupiter could cross the positive $x$-axis upwards at the same time, denoted by $t^{\ast}$. At this epoch, the FFP is located at its perihelion with $f=0$. Utilising the angular momentum integral of the two-body system comprising the Sun and the FFP
\begin{equation}
r^2\dot{f}=\sqrt{\tilde{\mu} p},
\end{equation}
and the transformation of Eq. (\ref{radial}), we can obtain
\begin{equation}
\frac{1}{4} p^2 \sec^4\frac{f}{2}df=\sqrt{\tilde{\mu} p}~dt.
\label{dfdt}
\end{equation}
Integrating both sides of Eq. (\ref{dfdt}) yields
\begin{equation}
\frac{1}{2}\tan\frac{f}{2}+\frac{1}{6}\tan^3\frac{f}{2}=n(t-t^{\ast}),
\label{fnt}
\end{equation}
where $n$ is the FFP's mean motion given by
\begin{equation}
n=\sqrt{{\tilde{\mu}}/{p^3}},
\label{nFFP}
\end{equation}
and the integral constant $t^{\ast}$ is the time of perihelion passage for the FFP as we defined above. By taking the initial true anomaly $f(0)$ of the FFP, we can eventually get
\begin{equation}
t^{\ast}=\left[\frac{1}{2}\tan\frac{f(0)}{2}+\frac{1}{6}\tan^3\frac{f(0)}{2} \right]/(-n).
\label{FFPperi}
\end{equation}
Requiring that Jupiter should also reach the position $(a_J^{\ast}, 0)$ at this very moment of $t=t^{\ast}$, we can reversely calculate the initial position angle of Jupiter to be
\begin{equation}
\theta^{\ast}=\frac{t^{\ast}\pmod{T_J}}{T_J}\cdot(-2\pi),
\label{sita}
\end{equation}
where $T_J\approx11.86$ yr is the orbital period of Jupiter with $a_J=a_J^{\ast}=5.2$ AU.
 
It should be noted that, the FFP has its orbit precessing when it comes close to Jupiter due to the mutual interaction, thus the position and time of perihelion passage are actually different from $(p/2, 0)$ and $t^{\ast}$ respectively, but just by very small amounts. So the analytical estimation made above for the time $t\le t^{\ast}$ is still applicable and will be used to provide initial conditions of these two planets in what follows.

\subsection{Results for jumping Jupiter}

\begin{figure}
  \centering
  \begin{minipage}[c]{0.5\textwidth}
  \centering
  \hspace{0cm}
  \includegraphics[width=8.5cm]{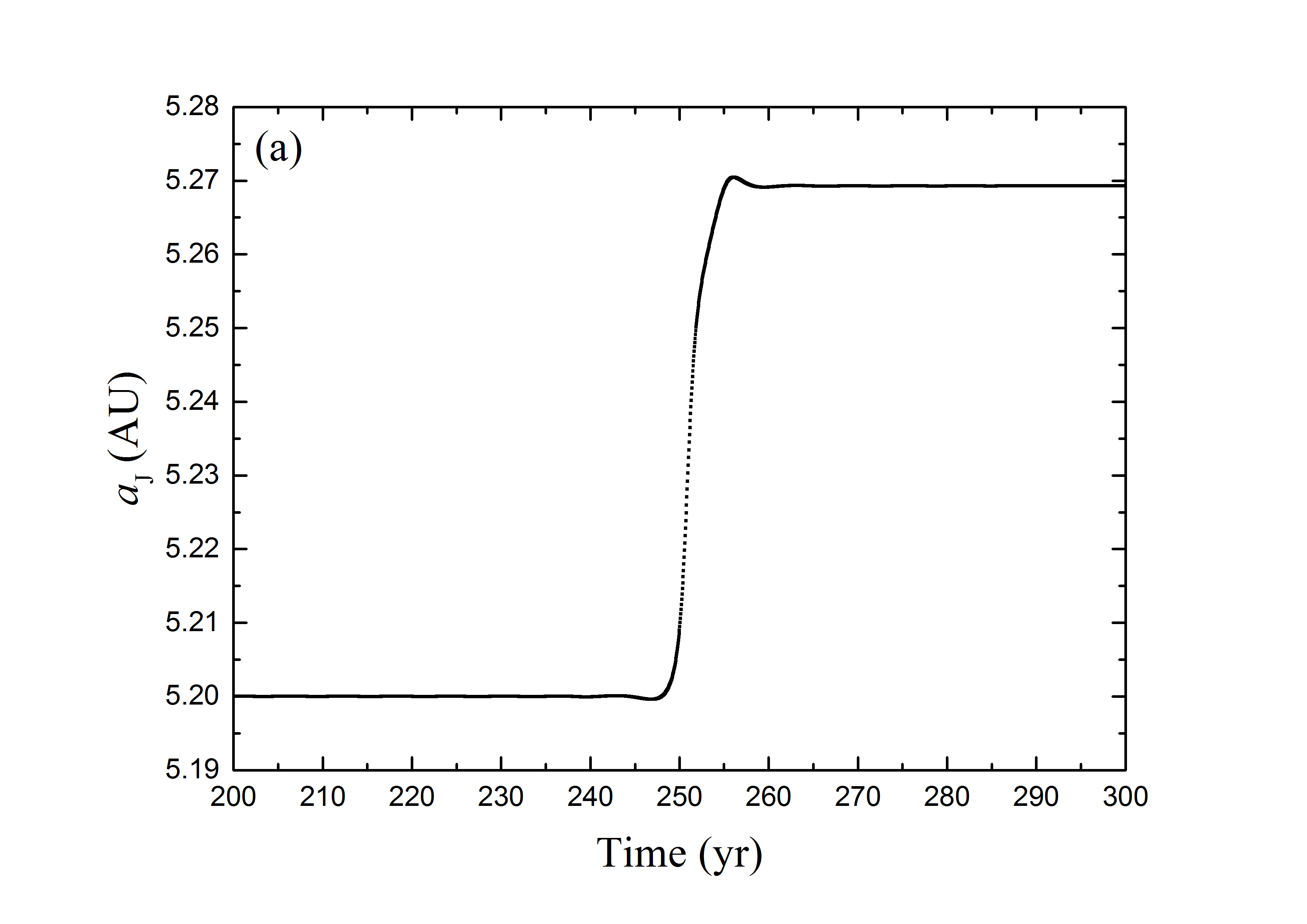}
  \end{minipage}
  \begin{minipage}[c]{0.5\textwidth}
  \centering
  \hspace{0cm}
  \includegraphics[width=8.5cm]{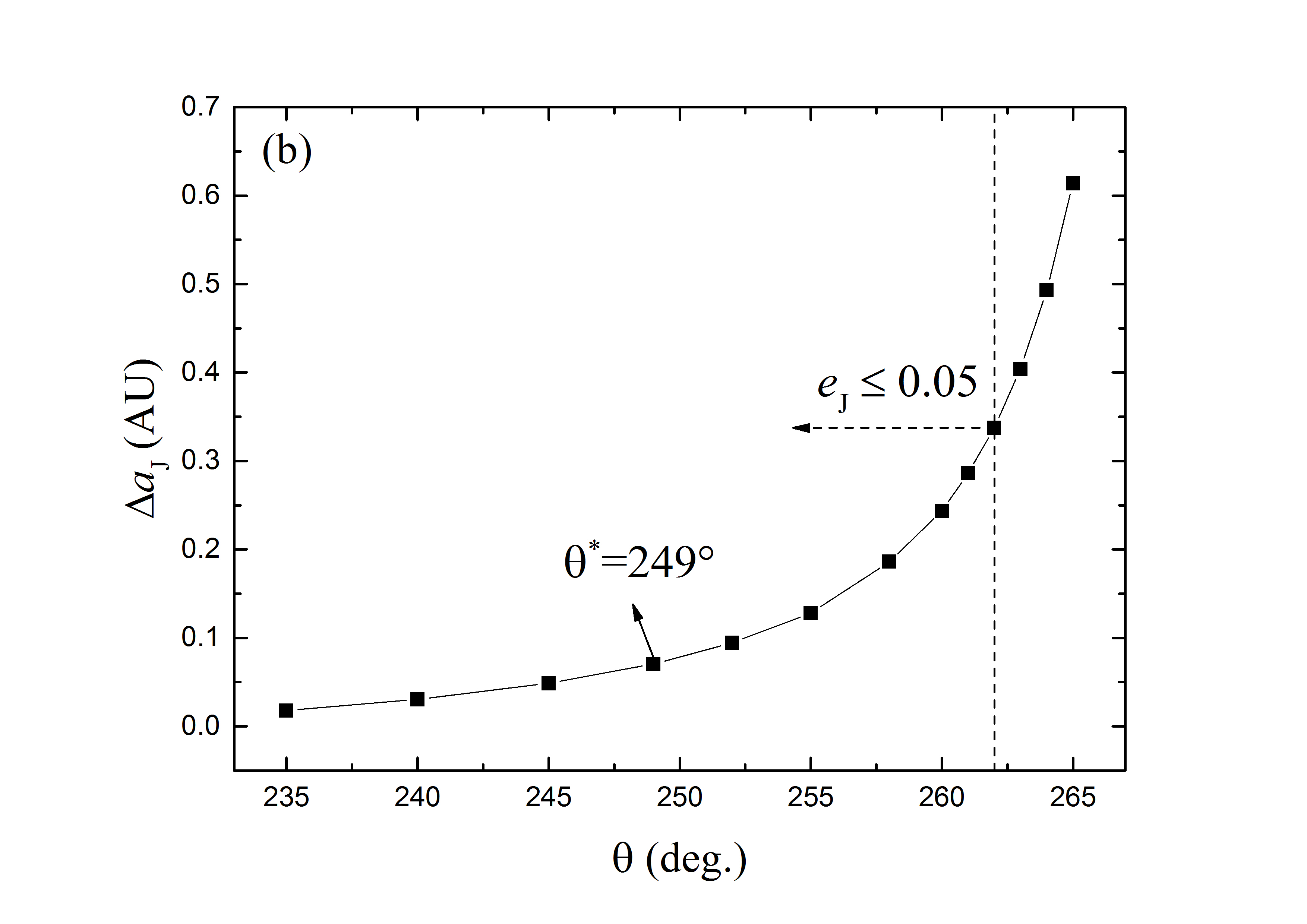}
  \end{minipage}
    \caption{An invading FFP with a distance parameter $d=13a_J^{\ast}$ and a mass $m_{FFP}=m_J$. Panel (a): temporal evolution of Jupiter's semimajor axis $a_J$, given $\theta=\theta^{\ast}=249^{\circ}$ (see Eq. (\ref{sita})). The sudden increase of $a_J$ happens just around the timing of the FFP's perihelion passage, that is $t^{\ast}=251$ yr. Panel (b): the migration amplitudes $\Delta a_J$ of Jupiter against different values of  $\theta$. An upper limit of $\Delta a_J$ is defined by the condition that Jupiter's eccentricity $e_J$ is smaller than its current value of $\sim0.05$.}
  \label{MigduetoFFP}
\end{figure}  

As we illustrated above, there are three free parameters to be considered: the distance $d$ measuring the initial $y$-coordinate of the FFP, the mass $m_{FFP}$ of the FFP and the initial position angle $\theta$ of Jupiter. After some tests, we have adopted $d=13a_J^{\ast}$ and $m_{FFP}=m_J$ for an invading FFP. Then Eq. (\ref{pvalue}) gives $p=2.06a_J^{\ast}$, indicating that this FFP has a perihelion at $(p/2, 0)=(1.03a_J^{\ast}, 0)$. Thus, it will be exterior to the orbit of Jupiter and exerts the perturbing acceleration on Jupiter's motion via a close encounter, as visualised in Fig. \ref{3body}. 

Since the orbit of the FFP is established, $\theta$ determines the minimum distance between the FFP and Jupiter, which is equivalent to the strength of their close encounter. In order to investigate the resulting jump of Jupiter, we have performed a series of numerical experiments for different values of $\theta$. It is quite obvious that the FFP only spends a small amount of time passing through the perihelion because of its high velocity. For instance, if we pick a fraction of the FFP's parabolic orbit where the radial distance is $r<2a_J^{\ast}$, the corresponding timespan is about 7 yr, as predicted by Eqs. (\ref{radial}) and (\ref{fnt}). Therefore, in this subsection, in order to accurately simulate the behaviour of Jupiter during FFP's pass-by, we employ the 7-8th-order Runge-Kutta-Fehlberg (RKF) algorithm with an adaptive stepsize and with a local truncation error of less than $10^{-15}$. The integration time is set to be 1000 yr, which is long enough compared to the time of the close encounter (i.e.  only of the order of a few years as we just estimated).


Fig. \ref{MigduetoFFP}(a) shows the typical temporal evolution of Jupiter's semimajor axis due to the brief visit of the FFP for $\theta=\theta^{\ast}=249^{\circ}$ as derived from Eq. (\ref{sita}). At the very beginning, Jupiter remains on its original orbit because the FFP is set to be coming from a distance as large as tens of $a_J^{\ast}$. Then, the FFP gradually approaches the orbit of Jupiter and Jupiter experiences a short period of outward migration on a timescale of about 10 yr (i.e. from $\sim246$ yr to $\sim256$ yr), as $a_J$ increases by about 0.07 AU. This timing of the close encounter is well consistent with the value $t^{\ast}=251$ yr calculated from Eq. (\ref{FFPperi}). After that, the FFP leaves the Solar System, and Jupiter switches to an expanded orbit, as indicated by the dashed circle in Fig. \ref{3body}. Here, the FFP does not notably excite Jupiter's eccentricity $e_J$, which is as small as about $<0.01$ at the end of the 1000 yr integration. 

Considering the fact that, due to the interaction between the two planets, the configuration of their close encounter would be somewhat different from what we designed theoretically. In other words, the strongest acceleration for Jupiter could take place at $\theta$-value other than $\theta^{\ast}$. Keeping the other initial conditions of the system unchanged, we repeat the 1000 yr numerical integrations by choosing $\theta$ to have a series of different values but close to $\theta^{\ast}$. The resultant migration amplitudes $\Delta a_J$ are plotted in Fig. \ref{MigduetoFFP}(b). The main feature is that within the adopted range of $\theta$, Jupiter could experience an even larger-scale migration with $\Delta a_J$ over 0.6 AU, while the migration timescale is still at the level of $\sim10$ yr. This way, the migration speed of Jupiter is nearly proportional to $\Delta a_J$.

Nevertheless, there are two possible limitations on the maximum value of $\Delta a_J$: (1) the eccentricity $e_J$ of Jupiter would be excited to a higher value along with the increase of $\Delta a_J$ when the FFP's perturbation becomes stronger. As shown in Fig. \ref{MigduetoFFP}(b), if we require the final $e_J$ to be below the currently observed value of $\sim0.05$, $\Delta a_J$ should not exceed 0.33 AU. (2) \citet{murr05} found that if Neptune migrates on a very short timescale, its asymmetric 2:1 resonance becomes unable to capture planetesimals at all. Similarly, if Jupiter migrates fast enough, its 1:1 resonance could not maintain the Trojans and all of them would escape. Therefore, a very large $\Delta a_J$ needs to be excluded, as we will show below.


By now, the perihelion of the FFP is fixed since the free parameter $d$ is adopted to be $13a_J^{\ast}$. Of course, a more distant encounter between the FFP and Jupiter is very likely when the FFP's perihelion increases. Bearing in mind that there is another free parameter, $m_{FFP}$, and  a more massive FFP could be introduced to counterbalance the weaker perturbation due to a larger relative distance for the encounter with Jupiter. Consequently, the acceleration for Jupiter's motion, measured by $\Delta a_J$, could still be comparable. A few additional tests with $d$ increasing together with $m_{FFP}$ have been carried out. We find that the $\Delta a_J$-profile always remains very similar with respect to that shown in Fig. \ref{MigduetoFFP}(b), while the $\theta$-range would slightly change since the timing of the FFP's perihelion passage is a little different. Taking the case of $d=15a_J^{\ast}$ and $m_{FFP}=10m_J$ for example, the perihelion of the FFP has been raised by about 2 AU, while the resulting $\Delta a_J$ can also reach quite a large value, up to 0.24 AU under the constraint of $e_J\lesssim0.05$.

Apart from the migration amplitude $\Delta a_J$, we also monitor the migration timescale, which is always at the level of 10 yr. It is easy to understand that by looking at the velocities of the FFP at perihelion (i.e. $f=0$):
\begin{equation}
\dot{x}_{peri}=0,~~~\dot{y}_{peri}=\sqrt{{\tilde{\mu}}/{p}}=\sqrt{\frac{{G(M_{\odot}+m_{FFP})}}{{\sqrt{(40a_J^{\ast})^2+d^2}-40a_J^{\ast}}}},
\end{equation}
where we substitute for $p$ from Eq. (\ref{pvalue}), and the subscript ``\textit{peri}'' represents the abbreviation for ``perihelion''. The FFP's mass $m_{FFP}$ has little effect on the velocity $\dot{y}_{peri}$ since the Sun's mass $M_{\odot}$ is significantly larger. Besides, when $d$ increases from $13a_J^{\ast}$ to $15a_J^{\ast}$, the variation of $\dot{y}_{peri}$ is only of the order of 10\%. Therefore, the time span of the perihelion traversing of the FFP would only change a little, as well as the duration of the close encounter between the FFP and Jupiter.

In summary, the invasion of a FFP could force Jupiter to jump outwards with a speed of $\dot{a}_J\approx{\Delta a_J}/$(10 yr), where ${\Delta a_J}$ can be as large as about 0.3 AU. Bearing in mind that, when $\dot{a}_J$ exceeds the value of $\dot{a}_J^{crit}=3.8\times10^{-3}$ AU/yr, in theory, the tadpole orbits around Jupiter's L4 point would disappear \citep{sica03, ogil06}. Thus, it is very interesting to see what would happen if Jupiter's migration speed exceeds that critical value, corresponding to ${\Delta a_J}>0.038$ AU for the 10 yr migration timescale found here. This is the fundamental difference between this work and \citet{li2023}.

\subsection{Results for L4/L5 asymmetry}

In this subsection, we study the evolution of the L4 and L5 Trojans for their number asymmetry in the jumping-Jupiter model. Same as in \citet{li2023}, the radial variation of Jupiter is mimicked by using the classical formula \citep{malh95}: 
\begin{equation}
  a_J(t)=a_J^{\ast}+\Delta a_J [1-\exp(-t/\tau)],
\label{eq:variation}
\end{equation}
where the e-folding time $\tau$ can represent the migration time $\Delta t$ according to the correlation 
\begin{equation}
\Delta t = (10/3)~\tau.
\label{taoTOdt}
\end{equation}
As for the test Trojans, they initially have semimajor axes $a=a_J^{\ast}=5.2$ AU, eccentricities $e=0$-0.3, inclinations $i=0.01$, while their longitude of ascending node $\Omega$ and argument of perihelion $\omega$ are chosen randomly between $0^{\circ}$ and $360^{\circ}$. Finally, the mean anomaly $M$ is determined from the resonant angle
\begin{equation}
  \sigma=(\Omega+\omega+M)-\lambda_J,
\label{eq:resangle}
\end{equation}
where $\lambda_J$ is the mean longitude of Jupiter. For clarity, we denote the initial resonant angle as $\sigma_0$. For the L4 and L5 swarms, we set $\sigma_0$ to be $60^{\circ}+\Delta \sigma_0$ and $-60^{\circ}-\Delta \sigma_0$ respectively, it means that these two populations are initially displaced from individual Lagrangian points by $\Delta\sigma_0$. The number $N_c$ of test Trojans for each swarm is adopted to be 500. 
More details can be found in Sect. 3.1 in \citet{li2023}.

\begin{table}
\centering
\begin{minipage}{8cm}
\caption{Statistics of surviving test Trojans within the model of Jupiter's outward migration. The migration rate is $\dot{a}_J=\Delta a_J/$(10 yr), where the possible amplitudes $\Delta a_J$ and the 10 yr timescale are obtained in the framework of the FFP encounter (see Sect. 2.2). Given the standard setting of $\Delta\sigma_0=30^{\circ}$-$80^{\circ}$, there are 500 test Trojans in each of the L4 and L5 swarms and the system is integrated for 1 Myr. We then derive the numbers ($N_4$, $N_5$) and final resonant amplitudes ($A_4$, $A_5$) of surviving Trojans (the subscripts `4' and `5' for the L4 and L5 swarms, respectively), as well as the number ratio $R_{45}=N_4/N_5$.}      
\label{model4}
\begin{tabular}{l| c c c c c}        
\hline                 
$\Delta a_J$ (AU)    &         $N_4$             &        $A_4$          &         $N_5$           &      $A_5$      & $R_{45}$                  \\

\hline\hline
           
0.11                           &          130                &        25-55             &         95              &        23-56       &        1.37           \\

\textbf{0.12}                         &      \textbf{122}               &        \textbf{25-52}             &         \textbf{74}             &        \textbf{35-55}       &        \textbf{1.65}           \\


0.15                      &                    65       &              35-55         &           25           &          43-55      &      2.60  \\


0.18                         &          20                &        44-52             &         3              &        42-54       &        6.67           \\

0.19                         &           13               &        48-55             &           0            &        --       &        --           \\


\hline
\end{tabular}
\end{minipage}
\end{table}

Regarding the system evolution, we consider three key parameters: the amplitude ${\Delta a_J}$ and the e-folding time $\tau$ for Jupiter's migration, together with the initial displacement $\Delta\sigma_0$ of the test Trojans from the associated Lagrangian point. According to the obtained jumping-Jupiter model induced by the FFP encounter, the timescale of 10 yr gives $\tau=3$ yr (see Eq. (\ref{taoTOdt})) and ${\Delta a_J}$ is allowed to vary in the range (0, 0.3] AU. For the test Trojans, we first choose samples having the standard $\Delta\sigma_0$ of $30^{\circ}$-$80^{\circ}$ as we did in \citet{li2023}, and there are 500 objects in each of the L4 and L5 swarms. Then we compute the orbital evolution of test Trojans by involving the jumping Jupiter with different ${\Delta a_J}$. To fulfill a long-term simulation, we employ the swift\_rmvs3 symplectic integrator developed by \citet{levi94}, with a time-step of 0.05 yr. At the end of the 1 Myr integration, the numbers of surviving L4 ($N_4$) and L5 ($N_5$) Trojans, the corresponding number ratio $R_{45}=N_4/N_5$, and their final resonant amplitudes $A_4$ and $A_5$ are reported in Table \ref{model4}.


We find that the L4/L5 number ratio $R_{45}$ increases very fast with ${\Delta a_J}$ (equivalent to the  migration speed). When ${\Delta a_J}\ge0.12$ AU, the resulting values of $R_{45}$ can be larger than 1.6, which is capable of explaining the current observations. More interestingly, at ${\Delta a_J}=0.19$ AU, $R_{45}$ finally becomes infinite since some L4 Trojans can remain stable, but all the L5 ones have escaped, that is, $N_4>0$ but $N_5=0$. Thus, $R_{45}$ could be as large as possible in the scenario of the FFP invasion. This is the major difference with \citet{li2023}. Besides, a further comparison with the results in \citet{li2023} raises two more differences: (i) when the migration of Jupiter is faster, the number $N_4$ of surviving L4 Trojans decreases; (ii) all the L4 Trojans have final resonant amplitudes larger than the original values deduced from the case of a non-migrating Jupiter. Both are in contrast with what we found in our previous work.
As a matter of fact, since the jumping Jupiter considered here having $\Delta a_J=0.12$-0.19 AU and $\Delta t=10$ yr, we immediately notice that its migration speed is beyond the critical value $\dot{a}_J^{crit}$  corresponding to the disappearance of the L4 librational islands. All these make us ask an intriguing question: unlike in \citet{li2023}, what is the peculiar evolution path that the L4 Trojans follow?

\begin{figure}
 \hspace{0 cm}
  \includegraphics[width=8.5cm]{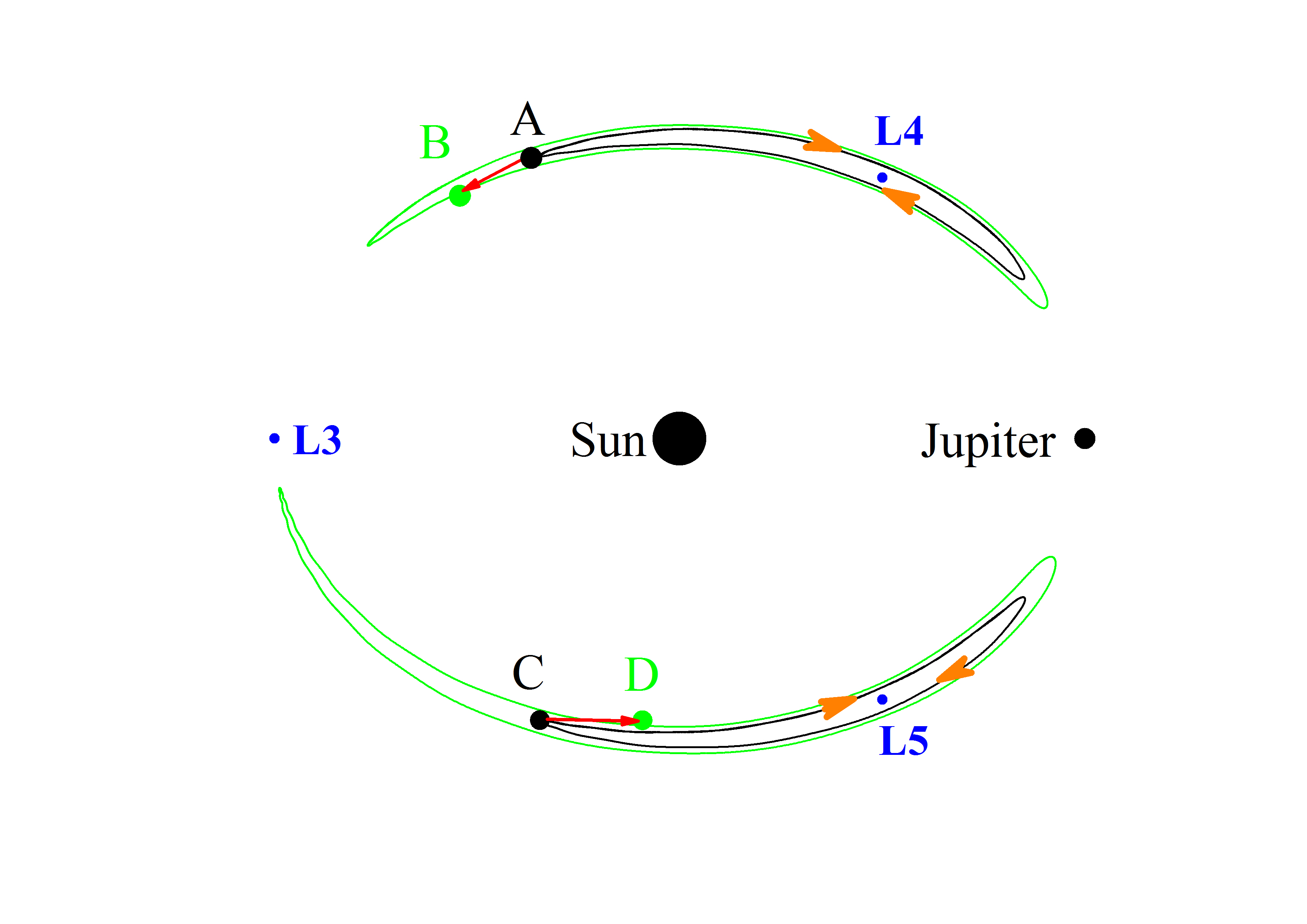}
  \caption{Schematic diagram of the evolution of Jupiter Trojans as a consequence of an outward migrating Jupiter with a speed $\dot{a}_J$ larger than $\dot{a}_J^{crit}$. This figure is plotted in a reference frame rotating with the mean motion of Jupiter. The L4 Trojan is assumed to start from point A. In the case of a non-migrating Jupiter, it will move on the original orbit (black curve), while in the case of a jumping-Jupiter with $\dot{a}_J>\dot{a}_J^{crit}$, it will first drift to point B and then move on the expanded orbit (green curve). Similarly, the L5 Trojan will switch from the black orbit to the green orbit after the drift from point C to point D caused by the jumping Jupiter. The direction of the tadpole motion is indicated by the orange arrow, for both the black and green orbits.}
  \label{SchemeDueToFFP}
\end{figure}

Fig. \ref{SchemeDueToFFP} sketches the tadpole orbits of the simulated Trojans in a frame rotating with the mean motion of Jupiter, before (black curves) and after (green curves) Jupiter's jump. The orange arrows indicate the directions of all the librational motions, for both the black and green orbits. A L4 Trojan has a starting position at point A. If there is no migration of Jupiter, this object would move along the black orbit towards L4 since it initially has the maximum $\sigma$ deviation. So the resonant angle $\sigma$ has to decrease at the beginning of the evolution. But if Jupiter migrates outwards at a speed higher than $\dot{a}_J^{crit}$ as we consider here, L4 will merge with L3 and the surrounding islands will disappear. As a result, the original L4 Trojans could temporarily switch to the horseshoe orbits and move towards the location of L3 \citep{ogil06}. As illustrate in Fig. \ref{SchemeDueToFFP}, such L4 Trojan, starting from point A, first goes to point B and its resonant angle $\sigma$ increases accordingly. Because the duration of this period is very short, equivalent to the 10 yr migration timescale of Jupiter, the angular distance between A and B would not be very large, that is, the increase of $\sigma$ should be a limited value. When the migration of Jupiter has ceased, shortly after, the L4 libration islands will reappear and this Trojan located at point B finally settles on the green tadpole orbit which possesses a larger resonant amplitude than that of the original black one.

\begin{figure}
  \centering
  \begin{minipage}[c]{0.5\textwidth}
  \centering
  \hspace{0cm}
  \includegraphics[width=8.5cm]{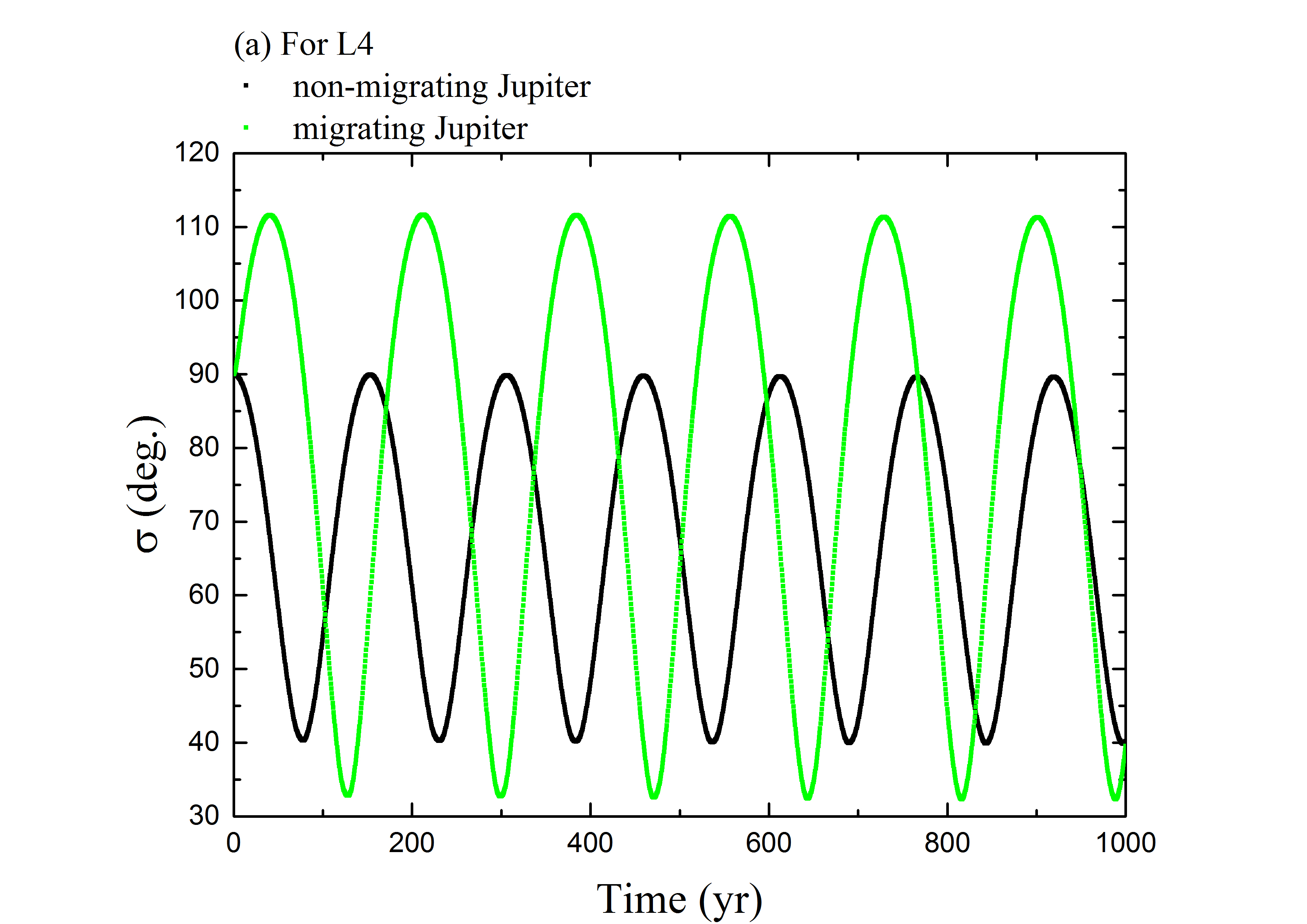}
  \end{minipage}
  \begin{minipage}[c]{0.5\textwidth}
  \centering
  \hspace{0cm}
  \includegraphics[width=8.5cm]{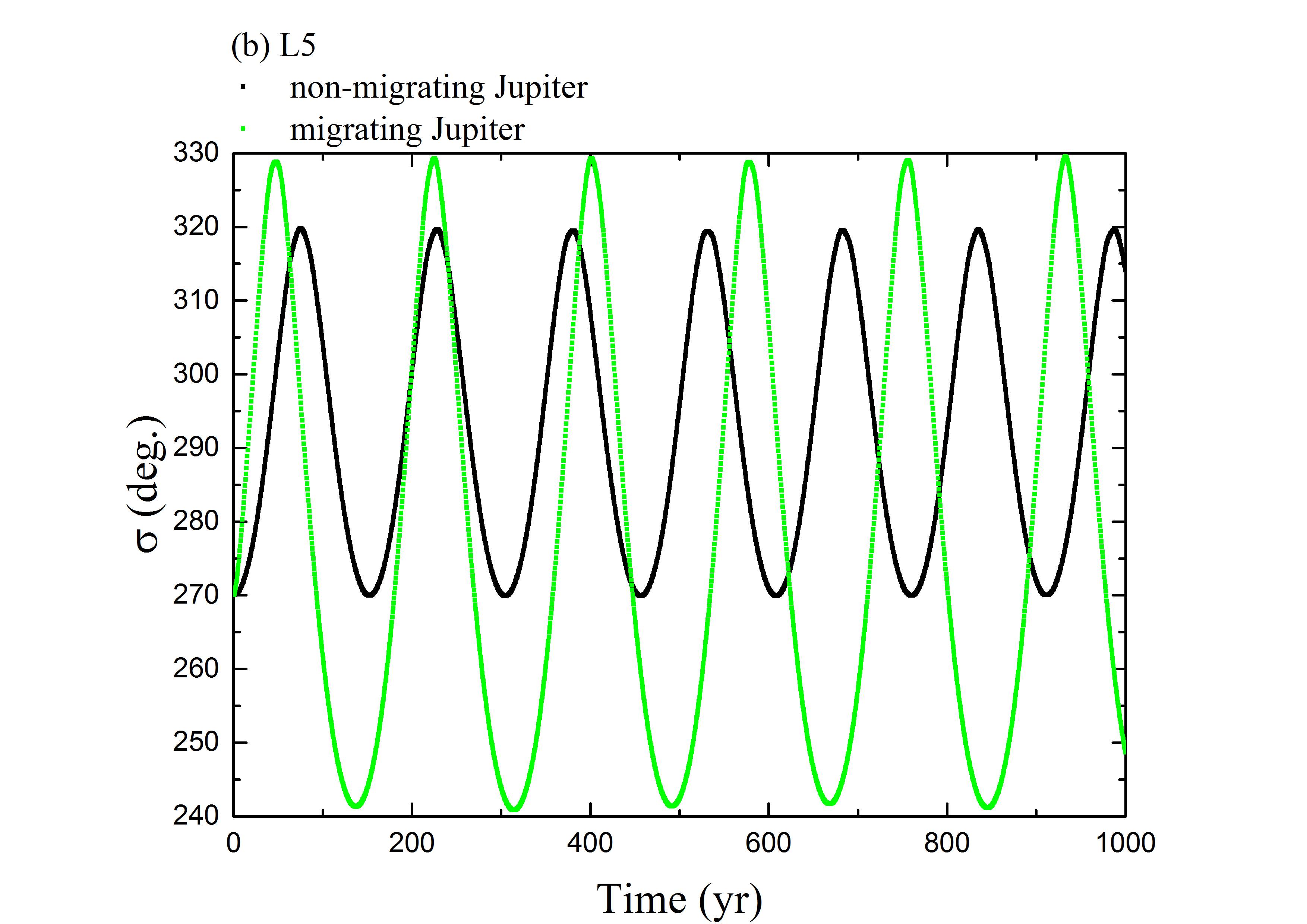}
  \end{minipage}
    \caption{Time evolution of the resonant angle $\sigma$ for the L4 (panel (a)) and L5 (panel (b)) test Trojans starting symmetric tadpole orbits, given initial $\sigma=90^{\circ}$ and $\sigma=270^{\circ}$ (i.e. $\sigma=-90^{\circ}$) respectively. In the case of a non-migrating Jupiter, the $\sigma$ evolution of these two objects are nicely symmetric, as indicated by the black curves.  In the case of an outward migrating Jupiter with speed $\dot{a}_J=0.015$ AU/yr ($>\dot{a}_J^{crit}$), as indicated by the green curves, both Trojans have their resonant amplitudes increased but the L4 one's resonant amplitude is less excited.}
  \label{SigmaDueToFFP}
\end{figure}

Corresponding to the schematic picture of Fig. \ref{SchemeDueToFFP}, we present in Fig. \ref{SigmaDueToFFP} the tadpole orbits generated by direct numerical simulations for different evolutional scenarios of Jupiter. The black curve refers to the case in which no migration is imposed on Jupiter and the green curve refers to the jumping-Jupiter model by assuming $\Delta a_J=0.15$ AU (i.e. $\dot{a}_J=0.015$ AU/yr $>\dot{a}_J^{crit}$). Panel (a) shows the temporal variations of $\sigma$ for the same L4 Trojan starting with $\Delta \sigma_0=30^{\circ}$. If there is no migration, the resonant angle $\sigma$ decreases immediately at the beginning of the integration. But in the migration case, $\sigma$ initially increases as this Trojan first moves towards the L3 point. At time of about 40 yr, that is, well after the 10 yr period of Jupiter's migration, $\sigma$ stops increasing because the Trojan is again restricted to the new L4 tadpole island with larger size (see the green orbit in Fig. \ref{SchemeDueToFFP}). As a result, the resonant amplitude is increased, from $25^{\circ}$ to $35^{\circ}$, as measured from the black and green orbits in Fig. \ref{SigmaDueToFFP}(a). This explains that, as presented in Table \ref{model4}, the minimum $A_4$ is about $35^{\circ}$ for the case of $\Delta a_J=0.15$. Similarly, the L4 Trojans with $\Delta \sigma_0>30^{\circ}$ will also experience the $A_4$ enlargement and some of them may not survive eventually. 




As Jupiter's migration rate $\dot{a}_J (>\dot{a}_J^{crit})$ increases, the Trojans around L4 could drift towards L3 more quickly during the temporary absence of the L4 islands. This implies that, in Fig. \ref{SchemeDueToFFP}, the distance between points A and B becomes larger, leading to a much wider (green) tadpole orbit. Since the migration timescale of Jupiter is fixed, the migration speed is proportional to ${\Delta a_J}$. Therefore, as ${\Delta a_J}$ increases, the two trends in $N_4$ and $A_4$ (see Table \ref{model4}) can be understood from the following: (i) the effect of the jumping-Jupiter on the resonant amplitude enlargement for the L4 swarm is enhanced and all the local Trojans will change to the tadpole orbits with larger and larger $A_4$. (ii) More and more L4 Trojans would be ejected as a result of orbital instability caused by increased $A_4$ and thus the number $N_4$ of survivals becomes smaller. It is important to emphasise that, even when the L4 point disappears during the short period of Jupiter's jump, the local Trojans are still restricted on the co-orbital horseshoe orbits \citep{ogil06}. As we found in the numerical simulations, the surviving L4 Trojans actually have their semimajor axes increased from initial values of $a_J^{\ast}$ to final values of $a_J^{\ast}+\Delta a_J$, together with the same change of Jupiter's semimajor axis.

In the framework of Jupiter's outward migration induced by the FFP encounter, the migration rate could exceed $\dot{a}_J^{crit}$.
In this scenario, although the L4 swarm follows a totally different evolutionary path in comparison with what we observed previously in \citet{li2023}, the L5 swarm follows a similar one because the L5 Lagrangian point always persists and consequently the surrounding islands do not cease to exist. Nevertheless, since the migration rate of Jupiter is very high now, the L5 islands would expand considerably in width, as visualised by the black (original) and green (final) tadpole orbits around L5 point in Fig. \ref{SchemeDueToFFP}. This leads to quite large resonant amplitudes ($A_5$) of the surviving L5 Trojans (see Fig. \ref{SigmaDueToFFP}(b)). 

Although both the L4 and L5 Trojan populations have their resonant amplitudes increased and the instability could occur subsequently, the former has more survivors persistently. Thus, the resulting number ratio $R_{45}$ is always larger than 1. This is because of the fact that, given the same migration rate $\dot{a}_J$, the enlargement of the resonant amplitude is more effective for the L5 swarm than for the L4 one. As an example shown in Fig. \ref{SigmaDueToFFP}, a L4 and a L5 Trojans are chosen to start with the same initial angular distances of $\Delta\sigma_0=30^{\circ}$ from the respective Lagrangian points. If Jupiter does not migrate, they can evolve on exactly symmetric tadpole orbits (black curves) with the same resonant amplitudes. But due to the migration of Jupiter at a speed of 0.015 AU/yr ($>\dot{a}_J^{crit}=3.8\times10^{-3}$ AU/yr), these two Trojans would change to wider tadpole orbits indicated by the green curves. We can see that both the L4 and L5 Trojans have their resonant amplitudes increased, from $25^{\circ}$, to $40^{\circ}$ and $45^{\circ}$ respectively. Accordingly, the resonant amplitude $A_4$ is $5^{\circ}$ smaller than $A_5$.

\begin{figure}
 \hspace{0 cm}
  \includegraphics[width=8.5cm]{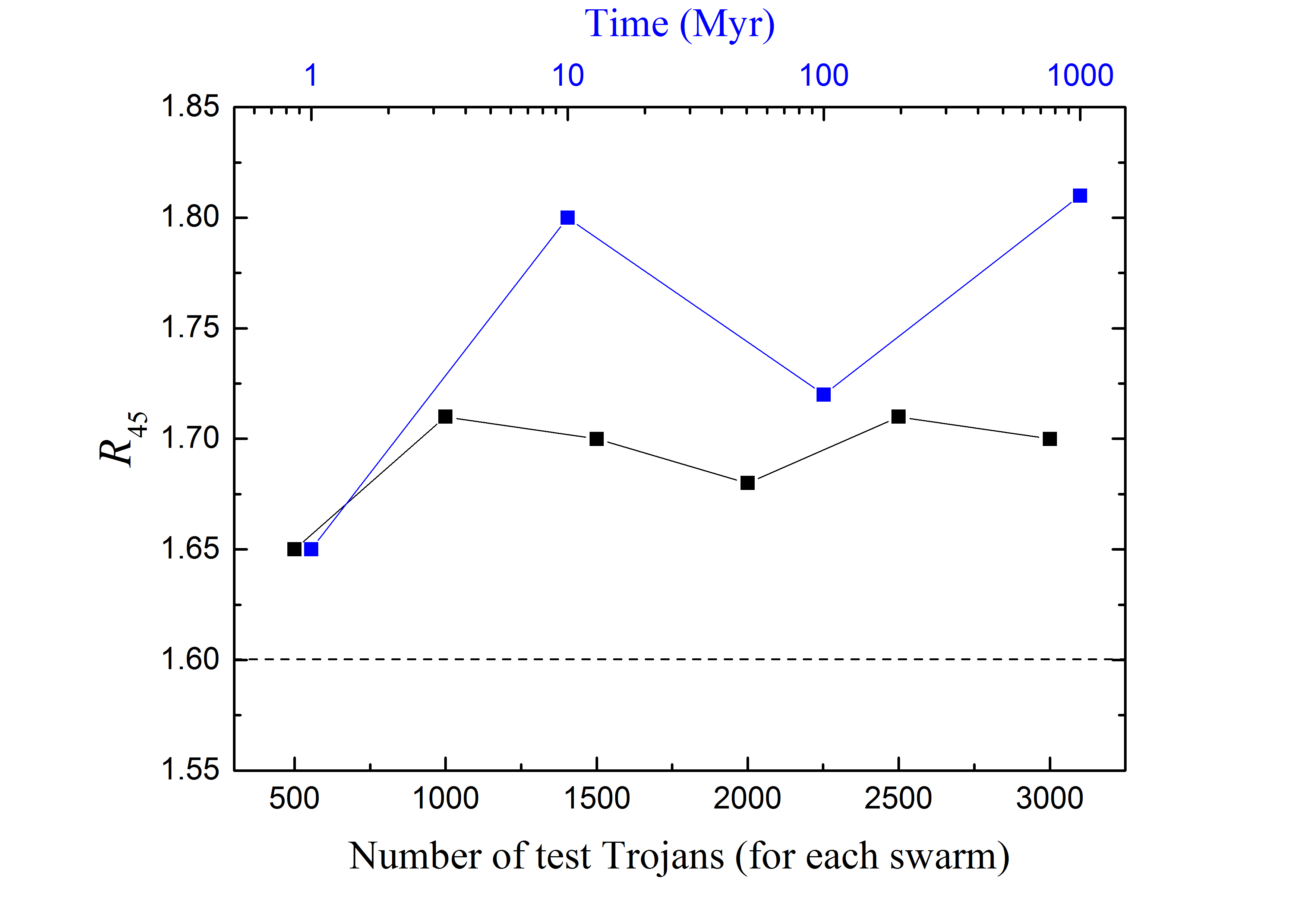}
  \caption{For the case of $\Delta a_J=0.12$ in Table \ref{model4} (highlighted in bold), the variation of the L4-to-L5 number ratio $R_{45}$ with larger numbers of test Trojans for each swarm (black curve) and longer integration times (blue curve). For reference, a horizontal dashed line is plotted at $R_{45}=1.6$, which corresponds to the unbiased number asymmetry of Jupiter Trojans.}
  \label{NandT}
\end{figure}

Before further proceeding, in relation to the determination of the reported number ratio $R_{45}$, we would like to provide justifications for our choices of: (1) the number $N_c=500$ of test Trojans for each swarm, and (2) the 1 Myr integration time. To investigate the effect of the number $N_c$ of test Trojans on $R_{45}$, for the case of $\Delta a_J=0.12$ in Table \ref{model4} (highlighted in bold), we performed 5 additional runs with the same parameters but distinct initial resonant angles $\sigma_0$ associated to the standard $\Delta \sigma_0$. As shown by the black curve in Fig. \ref{NandT}, when the number $N_c$ of test Trojans is increased from 500 to 3000 in increments of 500, the resulting $R_{45}$ values remain nearly constant in the narrow range of 1.65-1.7. This result indicates that, $N_c=500$ is sufficiently large to provide reliable statistics for the number asymmetry of surviving test Trojans. Next, we investigated the time dependence of the reported number asymmetry $R_{45}$, again for the case of $\Delta a_J=0.12$ in Table \ref{model4}. To do this, we extended the integration time to 1 Gyr, which is of the same order of magnitude as the age of the Solar System. The blue curve in Fig. \ref{NandT} shows that the number ratio $R_{45}$ varies with time, from 1.65 at 1 Myr to 1.81 at 1Gyr. This implies that the extended giga-year evolution may contribute an additional $\sim10$\% asymmetry, which is consistent with the previous studies \citep{disi14, li2023}. Thus we believe that the reported asymmetry is a permanent effect and could persist even if a longer integration time (e.g. 4.5 Gyr) is considered. Furthermore, it is noteworthy that larger $N_c$ and longer integration times may slightly alter $R_{45}$, but the resulting values are always greater than 1.6, as shown in Fig. \ref{NandT}. Based on this outcome, the current observation can surely be explained as we can simply reduce the perturbation on Jupiter, for example, by considering a FFP with smaller mass or larger perihelion. In a brief summary, to save computational time, in what follows we will use $N_c=500$ and an integration time of 1 Myr to estimate the leading-to-trailing number ratio $R_{45}$.

In \citet{li2023}, since the migration rate of Jupiter is limited to a value less than or comparable to $1.5\times10^{-4}$ AU/yr, which is derived by \citet{nesy13} from the era of the giant planet instability, we have to choose the standard $\Delta\sigma_0$ of $30^{\circ}$-$80^{\circ}$ in order to achieve the L4/L5 number ratio of $R_{45}\sim1.6$. But in this work, the migration of Jupiter could be much faster as long as the encounter between Jupiter and the FFP is close enough. Thus, for the same choice of $\Delta\sigma_0$, as listed in Table \ref{model4}, the number ratio $R_{45}$ can be considerably larger than 1.6. Based on this outcome, we suppose that if $\Delta\sigma_0$ has either smaller lower limit or smaller upper limit than that of the standard setting (i.e. the test Trojans start on more stable orbits closer to the Lagrangian points), the L4/L5 asymmetry may also reach the same level of $R_{45}\sim1.6$.

\begin{table}
\centering
\begin{minipage}{8.5cm}
\caption{The same as Table \ref{model4}, but for different choices of $\Delta\sigma_0$. As long as the migration of Jupiter is fast enough (i.e. $\Delta a_J$ is sufficiently large), the L4-to-L5 number ratio $R_{45}$ can always reach a value of $\sim1.6$, which is capable of explaining the current observation.}      
\label{FFPDiffSigma0}
\begin{tabular}{c c c c c c c}        

$\Delta\sigma_0=$$20^{\circ}$-$80^{\circ}$      \\
\hline\hline
$\Delta a_J$ (AU)    &         $N_4$             &       $A_4$          &         $N_5$           &      $A_5$      & $R_{45}$           \\
\hline


0.13                         &         143              &       26-50            &           106          &      28-51        &       1.35       \\

0.14                         &           126             &     29-52              &           79          &       32-53       &         1.59         \\\hline\hline\\

$\Delta\sigma_0=$$10^{\circ}$-$80^{\circ}$      \\
\hline\hline
$\Delta a_J$ (AU)    &         $N_4$             &       $A_4$          &         $N_5$           &      $A_5$      & $R_{45}$          \\
\hline

0.16                         &       117                 &       20-63            &       79              &     29-50         &   1.48                \\

0.17                         &       106                 &     28-51              &         63            &    37-53          &    1.68              \\\hline\hline\\

$\Delta\sigma_0=$$0$-$80^{\circ}$      \\
\hline\hline
$\Delta a_J$ (AU)    &         $N_4$             &       $A_4$          &         $N_5$           &      $A_5$      & $R_{45}$          \\
\hline
0.17                           &      120                  &    29-52               &       85              &    30-52          &        1.41       \\

0.18                         &          94              &      34-53             &          60           &      35-50        &        1.56           \\
\hline\hline\\

$\Delta\sigma_0=$$0$-$70^{\circ}$      \\
\hline\hline
$\Delta a_J$ (AU)    &         $N_4$             &       $A_4$          &         $N_5$           &      $A_5$      & $R_{45}$          \\
\hline
0.17                         &       130                 &          30-50     &      98              &       30-50        &        1.33           \\

0.18                           &           99              &        36-51            &           66           &        37-52       &      1.50        \\

\hline\hline\\

$\Delta\sigma_0=$$0$-$60^{\circ}$      \\
\hline\hline
$\Delta a_J$ (AU)    &         $N_4$             &       $A_4$          &         $N_5$           &      $A_5$      & $R_{45}$           \\
\hline

0.19                         &        79                &        36-51          &       60             &        36-52       &          1.32       \\

0.2                        &          67            &       41-56           &          41          &     43-52         &        1.63           \\

\hline\hline\\

$\Delta\sigma_0=$$0$-$50^{\circ}$      \\
\hline\hline
$\Delta a_J$ (AU)    &         $N_4$             &       $A_4$          &         $N_5$           &      $A_5$      & $R_{45}$           \\
\hline

0.19                         &          90              &     39-52              &         61            &    40-51          &          1.47         \\

0.2                         &            71            &       38-50            &         46            &     42-55         &       1.54            \\

\hline\hline\\

\end{tabular}
\end{minipage}
\end{table}

Similarly to what we did for obtaining Table \ref{model4}, by adopting different $\Delta\sigma_0$, we vary Jupiter's migration amplitude $\Delta a_J$ in order to see the resulting number difference between the L4 and L5 Trojans. At the end of the 1 Myr evolution, when $R_{45}$ reaches a value close to 1.6, we record the corresponding results, which are summarised in Table \ref{FFPDiffSigma0}. From the top to bottom, more and more test Trojans initially have rather small $\Delta\sigma_0$, equivalent to small initial resonant amplitudes. We find that, when Jupiter migrates fast enough by giving a large $\Delta a_J$, all of the test Trojans would have their resonant amplitudes significantly increased and a large fraction of them would escape from the Lagrangian regions. Since such a dynamical effect is more pronounced for the L5 swarm as analysed above, there would remain more Trojans populating the L4 region than the L5 one, and a number ratio $R_{45}\sim1.6$ can be achieved at some $\Delta a_J$. One may notice that, in the case of $\Delta\sigma_0=$$0$-$70^{\circ}$, given $\Delta a_J=0.18$ AU, $R_{45}=1.5$ seems a bit small. Actually, after the considered 1 Myr evolution, an extended Gyr evolution could further increase $R_{45}$ by a factor of $\sim1.1$, accounting for the current observed asymmetry \citep{li2023}.



As we can see in the last four cases shown in Table \ref{FFPDiffSigma0}, the larger the value of $\Delta a_J$, the larger the final resonant amplitudes $A_4$ and $A_5$. We would not consider test Trojans starting from even more stable orbits (e.g. with $\Delta\sigma_0=0$-$40^{\circ}$), because in that case an even large $\Delta a_J~(\gtrsim0.2~\mbox{AU})$ is needed, and accordingly all the surviving Trojans would end with resonant amplitudes $\gtrsim40^{\circ}$. We have to address that, by considering the jumping-Jupiter mechanism, the minimum resonant amplitudes of the simulated Trojans are over $15^{\circ}$-$20^{\circ}$, which are actually higher than those of the observed Trojans. Such an inconsistency was already discussed in \citet{li2023}, and a possible solution could be the collisions among Jupiter Trojans which may reduce their resonant amplitudes \citep{marz98, marz00}.


\subsection{Effects of FFP's parameters}

It is probable that the parameters of the FFP would affect the jumping-Jupiter scenario and the extent of the L4/L5 asymmetry would vary accordingly. In what follows, we evaluate the relative importance of the FFP's mass $m_{FFP}$ and orbital inclination $i_{FFP}$, which could be significantly different from what we adopted before (i.e. $m_{FFP}=m_J$ and $i_{FFP}=0^{\circ}$).

\subsubsection{FFP's mass}

\begin{figure}
 \hspace{0 cm}
  \includegraphics[width=9cm]{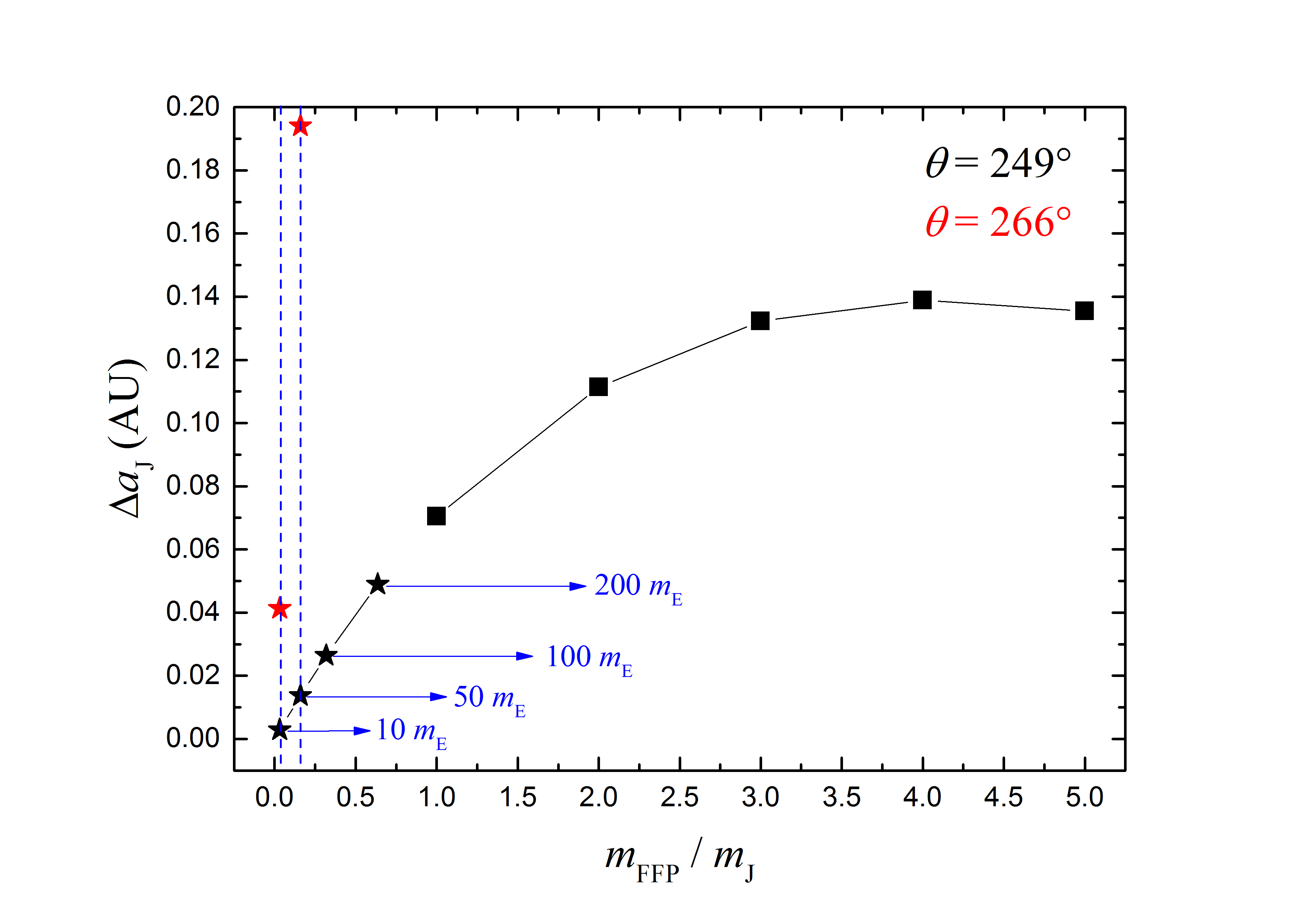}
  \caption{The migration amplitude $\Delta a_J$ of Jupiter as a function of the FFP's mass $m_{FFP}$ in the scenario of the close encounter. The values of $m_{FFP}$ are given in units of Jupiter masses $m_J$. For reference, less massive FFPs are indicated by their mass values in units of Earth masses $m_E$. The phase angles $\theta$ of Jupiter are taken to be: (i) the specific value of $\theta^{\ast}=249^{\circ}$ (black symbols); (ii) another representative value of $266^{\circ}$ (red symbols). The third parameter of the system, $d$, is the same as before ($d=13a_J^{\ast}$).}
  \label{FFPmass}
\end{figure}


An important contributor to Jupiter's migration amplitude $\Delta a_J$ is the mass $m_{FFP}$ of the passing-by FFP. Although $m_{FFP}$ is taken to be of the order of $m_J$ in the literature \citep{varv12, park17, goul18}, as well as in our previous model, this mass value may vary within quite a large range. Indeed, FFPs could have masses up to about 13 Jupiter masses and down to only a few Earth masses \citep{meli08}. 

Similar to the model constructed for the process of Jupiter jumping as described in Sect. 2.2, we perform some extra runs by choosing different values of $m_{FFP}$ in the range of $10~m_E$ to $5~m_J$, where the Earth mass $m_E$ is about $0.003~m_J$. The other two parameters $d=13a_J^{\ast}$ and $\theta=\theta^{\ast}=249^{\circ}$ are kept the same. The results are summarised in Fig. \ref{FFPmass}, as shown by the black symbols. From left to right, the four stars indicate the cases of $m_{FFP}=10$, 50, 100, and 200$~m_E$ and the four squares indicate the more massive FFPs with $m_{FFP}=1$-$5~m_J$. We can see that, for $m_{FFP}\le 1~m_J$, the migration amplitude of Jupiter depends nearly linearly on the FFP's mass, that is, $\Delta a_J\propto m_{FFP}$. This is easy to understand that the Newtonian gravitational acceleration in the motion of Jupiter is proportional to the mass of the perturber FFP. We further note that, when $m_{FFP}$ is equal to or greater than $2~m_J$, the value of $\Delta a_J$ seems not to follow the linear increase anymore. This is because the much stronger perturbation of such massive FFP would induce very different geometric configuration of the Jupiter-FFP encounter (e.g. the minimum relative distance). However, according to Table \ref{model4}, only the most massive FFPs with $m_{FFP}\gtrsim2~m_J$ can produce $\Delta a_J\gtrsim0.12$, leading to the required L4/L5 asymmetry of $R_{45}\sim1.6$.

We now take a look in Fig. \ref{FFPmass} at the cases of the smallest masses of $m_{FFP}=10$ and 50$~m_E$, which are highlighted by the vertical dashed lines. For the considered $\theta=249^{\circ}$, we find the resulting $\Delta a_J$ to be only about $<0.02$ AU (see the two leftmost black stars), which seems too small to generate the current L4/L5 asymmetry as Table 1 suggests that $\Delta a_J\gtrsim0.12$ AU is needed. Bear in mind that, the strongest perturbation of the FFP happens for Jupiter with the initial phase angle $\theta$ somewhat larger than $249^{\circ}$, as shown in Fig. \ref{MigduetoFFP}. When we adjust $\theta$ to be $266^{\circ}$ and redo the simulations, we find that, for $m_{FFP}=10$ and 50$~m_E$, the values of $\Delta a_J$ increase to 0.04 and 0.19 AU respectively (see the red stars in Fig. \ref{FFPmass}). Accordingly, we propose that $\Delta a_J$ could always reach a value of $>0.19$ AU for $m_{FFP}$ from $\sim50~m_E$ to a few $m_J$, since the gravitational effect of the FFP on Jupiter is stronger at larger $m_{FFP}$. We want to note that, in practice, it is meaningless to consider the cases of $\theta>266^{\circ}$, as the motion of Jupiter could become  chaotic due to the extremely close encounter between Jupiter and the FFP.

At this point, we conclude that a sub-Saturn or Jupiter-like FFP (i.e. with tens to hundreds of times the mass of Earth) can induce $\Delta a_J\gtrsim0.12$ AU that can explain the L4/L5 number asymmetry of Jupiter Trojans, but the less massive super-Earth FFP (i.e. with a few Earth masses) may fail to do so. In addition, we have to point out that in all these cases, Jupiter ends on a nearly circular orbit with eccentricity $e_J<0.05$.

\subsubsection{FFP's inclination}

So far, we have considered the FFP travelling on a parabolic orbit in the same plane as Jupiter's orbit, near the invariable plane of the Solar System \citep{li2019}. It is, however, more likely that the FFP penetrated the Solar System on an inclined orbit. 

By looking at Fig. \ref{3body}, in order to incline the orbital plane of the FFP, we can carry out a rotation about the $y$-axis through a positive angle $i_{FFP}$. Such transformation can be represented by introducing the $3\times3$ rotation matrix
\begin{equation}
\mathcal{R}=\left(\begin{array}{ccc}
\cos{i_{FFP}}  &  0  & \sin{i_{FFP}} \\
      0        &  1  &      0         \\
-\sin{i_{FFP}} &  0  &  \cos{i_{FFP}}
\end{array}\right)
\label{rotate}
\end{equation}
and consequently the initial conditions of the FFP given in Eq. (\ref{initialFFP}) are modified as
\begin{equation}
\left(\begin{array}{c}x(0) \\ y(0) \\  z(0)  \end{array}\right)
=\mathcal{R}\left(\begin{array}{c}-40a_J^{\ast} \\ -d \\  0  \end{array}\right)
\label{incliedX}
\end{equation}
and
\begin{equation}
\left(\begin{array}{c}\dot{x}(0) \\ \dot{y}(0) \\  \dot{z}(0)  \end{array}\right)
=\mathcal{R}\left(\begin{array}{c}-\sqrt{{\tilde{\mu}}/{p}}\sin f(0) \\ \sqrt{{\tilde{\mu}}/{p}}~(1+\cos f(0)) \\  0  \end{array}\right).
\label{incliedV}
\end{equation}

Fig. \ref{3body} shows that the minimum distance between Jupiter and the FFP occurs at the location of the FFP's perihelion on the positive $x$-axis. One can easily picture that, when the FFP's inclination increases by rotating its orbit about the $y$-axis, the FFP's perihelion moves away from Jupiter's orbital plane, and accordingly the minimum distance between these two planets becomes larger. As a result, the gravitational acceleration on Jupiter caused by the FFP is weaker and the migration amplitude $\Delta a_J$ should be smaller. However, the duration of the perihelion passage of the FFP can hardly change and thus the timescale of the close encounter between Jupiter and the FFP would be still around 10 yr. Accordingly, Jupiter's outward migration slows down, inducing less prominent number asymmetry of the L4 and L5 Trojans.

When the FFP's orbital plane shifts as the inclination $i_{FFP}$ increases, the $z$-coordinate of its perihelion is given by 
\begin{equation}
z^p_{FFP}=-p/2\cdot\sin{i_{FFP}},
\label{periZ}
\end{equation}
where the semilatus rectum $p$ of the FFP's parabolic orbit is adopted to be $2.06a_J^{\ast}$ as before. It means that the minimum distance between Jupiter and the FFP's perihelion would not be smaller than $z^p_{FFP}$. We then think that, if the orbital plane of the FFP is elevated along the positive $z$-axis by a magnitude of $\Delta z^p_{FFP}$, the weakened perturbation on Jupiter may be compensated. 


\begin{table}
\centering
\hspace{-2cm}
\begin{minipage}{7.5cm}
\caption{The excitation of Jupiter's orbit (i.e. the migration amplitude $\Delta a_J$, the final eccentricity $e_J$ and inclination $i_J$) within the context of a FFP approaching Jupiter. The FFP travels on a parabolic orbit with an inclination of $i_{FFP}$ relative to the orbital plane of Jupiter, and its initial conditions are given by Eqs. (\ref{incliedX}) and (\ref{incliedV}). The parameter $z^p_{FFP}$ indicates the vertical displacement of the FFP's perihelion from Jupiter's orbital plane ($z=0$), and it increases with $i_{FFP}$. At each $i_{FFP}$, the FFP's orbit is allowed to be elevated along the positive-$z$ axis by a magnitude of $\Delta z^p_{FFP}$. This way, the perihelion of the inclined FFP could be closer to Jupiter.}      
\label{FFPinc}
\begin{tabular}{c c c c}        

$i_{FFP}=0^{\circ}$~~~($z^p_{FFP}=0$)      \\
\hline\hline
$\Delta z^p_{FFP}$ ($a_J^{\ast}$)  & $\Delta a_J$ (AU)   &  $e_J$  &  $i_J(^{\circ})$   \\
\hline

0                         &         0.24        &  0.037  &  0     \\
\hline\hline\\

$i_{FFP}=10^{\circ}$~~~($z^p_{FFP}=-0.18a_J^{\ast}$)      \\
\hline\hline
$\Delta z^p_{FFP}$ ($a_J^{\ast}$)  & $\Delta a_J$ (AU)   &  $e_J$  &  $i_J(^{\circ})$   \\
\hline

0                         &    0.14                &  0.020    &  0.5      \\
\hline
0.1                       &    0.23                &  0.034    &  0.3      \\

\hline\hline\\

$i_{FFP}=20^{\circ}$~~~($z^p_{FFP}=-0.35a_J^{\ast}$)      \\
\hline\hline
$\Delta z^p_{FFP}$ ($a_J^{\ast}$)  & $\Delta a_J$ (AU)   &  $e_J$  &  $i_J(^{\circ})$   \\
\hline

0                         &         0.07          &  0.008    &  0.4     \\
\hline
0.1                       &         0.11          &  0.014    &  0.5      \\
\hline
0.2                       &         0.19          &  0.027    &  0.5     \\

\hline\hline\\

$i_{FFP}=30^{\circ}$~~~($z^p_{FFP}=-0.52a_J^{\ast}$)      \\
\hline\hline
$\Delta z^p_{FFP}$ ($a_J^{\ast}$)  & $\Delta a_J$ (AU)   &  $e_J$  &  $i_J(^{\circ})$   \\
\hline

0                         &         0.04          &  0.004   &  0.2     \\
\hline
0.2                       &         0.08          &  0.010   &  0.4     \\
\hline
0.3                       &         0.14          &  0.019   &  0.6     \\
\hline
0.4                       &         0.30          &  0.045   &  0.6    \\

\hline\hline\\

$i_{FFP}=40^{\circ}$~~~($z^p_{FFP}=-0.66a_J^{\ast}$)      \\
\hline\hline
$\Delta z^p_{FFP}$ ($a_J^{\ast}$)  & $\Delta a_J$ (AU)   &  $e_J$  &  $i_J(^{\circ})$   \\
\hline

0                         &         0.02          &  0.002   &  0.2     \\
\hline
0.3                       &         0.06          &  0.006   &  0.4     \\
\hline
0.4                       &         0.10          &  0.011   &  0.7   \\
\hline
0.5                       &         0.27          &  0.039   &  1.2    \\

\hline\hline\\

$i_{FFP}=50^{\circ}$~~~($z^p_{FFP}=-0.79a_J^{\ast}$)      \\
\hline\hline
$\Delta z^p_{FFP}$ ($a_J^{\ast}$)  & $\Delta a_J$ (AU)   &  $e_J$  &  $i_J(^{\circ})$   \\
\hline

0                         &         0.02           &  0.002   &  0.1      \\
\hline
0.4                       &         0.04          &   0.002   &  0.4      \\
\hline
0.5                       &         0.05          &   0.003    &  0.8     \\
\hline
0.6                       &         0.04          &   0.011    &  2.3   \\
\hline
0.7                       &     \mbox{chaotic}    &   -        & -    \\

\hline\hline\\

\end{tabular}
\end{minipage}
\end{table}

The results of our simulations with different values of $i_{FFP}$ are presented in Table \ref{FFPinc}. Since the timing of the close encounter is similar to the planar case, according to Fig. \ref{MigduetoFFP}, we choose the phase angle of Jupiter to be $\theta=260^{\circ}$, which can amplify the migration amplitude $\Delta a_J$ for Jupiter but would not induce chaotic evolution. For the FFP moving on the original orbit starting with the positions and velocities provided by Eqs. (\ref{incliedX}) and (\ref{incliedV}) (i.e. for $\Delta z^p_{FFP}=0$), we can see that $\Delta a_J$ decreases monotonically with larger $i_{FFP}$. At $i_{FFP}\ge20^{\circ}$, $\Delta a_J$ goes down to a value below 0.12 AU, which serves as the lower limit of the migration amplitude of Jupiter that could produce the Trojan number asymmetry of $R_{45}\gtrsim1.6$ according to the results shown in Table \ref{model4}. 

When the FFP's orbital plane is moving upwards along the $z$-axis, that is for $\Delta z^p_{FFP}>0$, we find that $\Delta a_J$ would indeed become larger (see Table \ref{FFPinc}). For instance, at $i_{FFP}=40^{\circ}$, we have $\Delta a_J=0.10$ AU and 0.27 AU when $\Delta z^p_{FFP}$ is set to be $0.4a_J^{\ast}$ and $0.5a_J^{\ast}$ respectively. Thus, a proper $\Delta z^p_{FFP}$ between these two values can be found to give $\Delta a_J\gtrsim0.12$, to account for the number ratio $R_{45}\sim1.6$ of Jupiter Trojans. We would like to note that, when $\Delta z^p_{FFP}=0$, the minimum distance between Jupiter and the FFP is about $0.53a_J^{\ast}$. Such a large mutual distance indicates that the gravitational interaction would be extremely small, characterised by $\Delta a_J=0.02$ AU. But when $\Delta z^p_{FFP}=0.5a_J^{\ast}$, 
we find that the said minimum relative distance decreases to only about $0.08a_J^{\ast}$, and consequently Jupiter's migration amplitude $\Delta a_J$ can much larger, of 0.27 AU. Therefore, the required strength of the FFP's perturbation can be well met in this circumstance.


However, for a more inclined FFP with $i_{FFP}=50^{\circ}$, there seems to be no significant increase in $\Delta a_J$ when $\Delta z^p_{FFP}$ increases from 0 to $0.6a_J^{\ast}$, as $\Delta a_J$ is always of the order of $\le0.05$ AU. This could be due to the fact that, when $i_{FFP}$ exceeds $45^{\circ}$, the acceleration caused by the FFP along the velocity of Jupiter in the $x-y$ plane becomes smaller than that along the $z$ direction. In addition, for an even larger value of $\Delta z^p_{FFP}=0.7a_J^{\ast}$, the relative distance between Jupiter and the FFP could reduce to be so small that Jupiter would experience the chaotic evolution because of an extremely close encounter with the FFP.

In summary, in order to explain the unbiased L4/L5 number asymmetry of $R_{45}\sim1.6$, the FFP that invaded the Solar system should have an inclination with respect to Jupiter's orbital plane at the level of $\lesssim40^{\circ}$. We need to clarify that such an inclined FFP could excite Jupiter's orbital eccentricity $e_J$ and inclination $i_J$, but which are always as small as $<0.04$ and $\lesssim2^{\circ}$, respectively, as shown in the last two columns in Table \ref{FFPinc}. According to our results in \citet{li2023}, such small $e_J$ and $i_J$ have little effect on the L4/L5 asymmetry in the jumping-Jupiter model.

\section{Direct case: FFP traversing the L5 region}

As we conjectured in the introduction, if a FFP once traversed the L5 region, it could deplete a number of local Trojans, while the L4 Trojans would not be affected due to large relative distances (i.e. at the level of the distance between the L4 and L5 point). Such a direct influence of the FFP on the Jupiter Trojans may also induce the L4/L5 number asymmetry. Therefore, we would like to investigate whether this mechanism could explain the required extent of the asymmetry of $R_{45}\sim1.6$.


\subsection{The control model}

In Sect. 2, in order to achieve the close encounter between Jupiter and the FFP, we determined the initial position phase $\theta$ of Jupiter to be $\theta^{\ast}$ (see Eq. (\ref{sita})). Accordingly, one can make the FFP approach the L5 point, of which Jupiter is $60^{\circ}$ ahead, by simply adopting $\theta=\theta^{\ast}+60^{\circ}$. Given the same mass and orbit of the FFP (i.e. $m_{FFP}=m_J$, $d=13a_J^{\ast}$), the deduced parameter $\theta^{\ast}=249^{\circ}$ implies the new $\theta=309^{\circ}$. In order to simulate accurately the perturbation of the FFP at the time of its perihelion passage, we employ again the RKF algorithm to integrate a system consisting of the Sun, Jupiter, a FFP and two swarms of test Trojans. For either the L4 or L5 swarm, there are 500 objects; the major difference is that, we choose a rather small $\Delta\sigma_0$ of $0$-$40^{\circ}$ according to the observed Jupiter Trojans.

As it was noted before, the motions of Jupiter and the FFP in numerical simulations could be somewhat different from our theoretical prediction due to their mutual interactions. Here, we find that  when the FFP arrives at its perihelion, the angular difference from Jupiter's location is about $40^{\circ}$, indicating that the L5 point is still $20^{\circ}$ behind the FFP. It suggests that we need to increase the angular position of the L5 point by $20^{\circ}$, equivalently, we increase Jupiter's initial position phase $\theta$ from $309^{\circ}$ to $330^{\circ}$. Under such a setting, the FFP can indeed come close to the L5 point around its perihelion and possibly scatter a large number of the local Trojans. We integrate the system for 300 yr, after which the FFP has already been far away from Jupiter's orbit as it reaches a heliocentric distance of $>13a_J^{\ast}$. Therefore, the FFP will never perturb Jupiter Trojans again on its way leaving the Solar System.  

\begin{figure}
 \hspace{0 cm}
  \includegraphics[width=8.5cm]{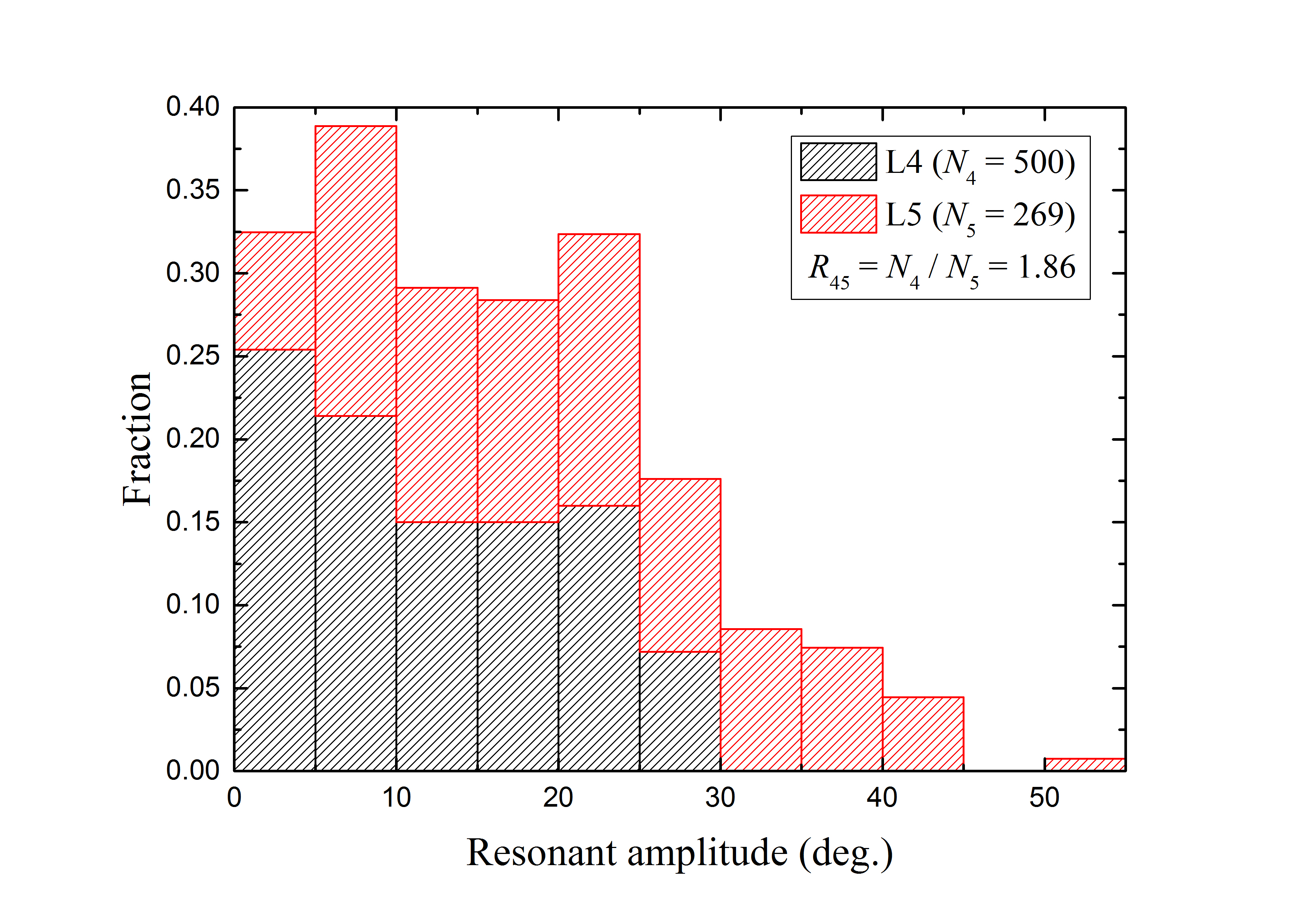}
  \caption{Distribution of resonant amplitudes for surviving L4 (black) and L5 (red) Trojans in the model of a Jupiter-mass FFP traversing the L5 region. The final L4-to-L5 number asymmetry can reach a value of $R_{45}=1.86$ (see legend within figure). Besides, a considerable fraction of these simulated Trojans have resonant amplitudes smaller than $30^{\circ}$, similar to the observed objects.}
  \label{FFPtraverse}
\end{figure}

In the subsequent evolution, only Jupiter and its Trojan population  orbit the Sun in the system. For the sake of saving computational time, we instead use the swift\_rmvs3 integrator to continue the system's evolution for another 1 Myr, without the FFP that has already left. At the end of the integration, we identify 269 L5 Trojans surviving on the tadpole orbits, while for the L4 swarm, since its population is always at great distances away from the FFP, all of the 500 members have survived. With this difference, we get a number ratio of $R_{45}=1.86$. Therefore, it is entirely plausible that the observed asymmetry of $R_{45}=1.6$ can be obtained if the perturbations of the FFP on the L5 Trojans are a bit weaker, for instance, in the case of a less massive FFP or a larger encounter distance.

For the Jupiter Trojans, besides explaining the current leading-to-trailing number ratio, the mechanism of a FFP invading the L5 region also works well in retaining the Trojans' small resonant amplitudes, as shown in Fig. \ref{FFPtraverse}. When the FFP penetrates deeply in the L5 region, not all of the local Trojans have the chance to experience close encounters with the FFP and some objects can remain on the unexcited tadpole orbits. These survivors may have their resonant amplitudes nearly unchanged, mainly below $30^{\circ}$ as the observation tells us. As for the L4 Trojans, surely they would keep the original resonant amplitude distribution due to the absence of strong perturbations from the FFP. Therefore, this new model could have an additional advantage of reproducing the current observed features of Jupiter Trojans. 

\subsection{Effects of the FFP's parameters}

As shown above, in the direct case, a FFP that traversed the L5 region once can well account for the number asymmetry of Jupiter Trojans, given its mass $m_{FFP}=m_J$ and inclination $i_{FFP}=0^{\circ}$ (i.e. in the same plane of Jupiter's orbit). Similar to what we did in Sect. 2, it is also of great significance to investigate whether such direct mechanism could be applied to a less massive or more inclined FFP.

\subsubsection{FFP's mass}

Considering the control model constructed in Sect. 3.1, we first reduce the FFP's mass $m_{FFP}$ from $1~m_J$ to $100~m_E$, while keeping other system parameters the same. Then we simulate again the process of FFP invading the L5 region. As reported in Table \ref{traveseMffp} (the first two rows), the L4 Trojans are still not affected at all, but there are much more L5 Trojans that could survive, leading to a decrease in the leading-to-trailing number ratio $R_{45}$ from 1.86 to 1.12. 

Bearing in mind that, besides $m_{FFP}$, the parameters $d$ and $\theta$ are also adjustable. In Sect. 2, we chose $d=13~a_J^{\ast}$ to avoid the collision between the FFP and Jupiter, since $d=12.8~a_J^{\ast}$ yields a FFP's perihelion located just on the orbit of Jupiter. But here we can adopt $d=12.8~a_J^{\ast}$ to let the FFP penetrate the centre of the L5 region and accordingly, we revise the value of $\theta$ which describes the phase angle of Jupiter as well as of its Trojan population. Table \ref{traveseMffp} (the last three rows) shows that, given $d=12.8~a_J^{\ast}$, the FFP with $m_{FFP} \ge 50~m_E$ is capable of depleting a sufficiently large number of L5 Trojans when it travels into this swarm and the resulting number ratio $R_{45}$ could reach the value of $>1.6$, to account for the current asymmetry problem. A much smaller FFP with $m_{FFP}= 10~m_E$, however, does not work no matter how we adjust the values of $d$ and $\theta$. It is quite interesting to point out that, similar to what we found in the jumping-Jupiter model (see Sect. 2.4), the mechanism of FFP invading the L5 Trojans also needs a sub-Saturn FFP at least.

\begin{table}
\centering
\begin{minipage}{8cm}
\caption{Statistics of surviving test Trojans in the model of FFP traversing the L5 region. The first row refers to the control model described in Sect. 3.1. In the following rows, we reduce the FFP's mass $m_{FFP}$, and the other two parameters $d$ and $\theta$ of the system are adjustable.}      
\label{traveseMffp}
\begin{tabular}{c c c | c c c}        
\hline                 
$m_{FFP}$ &   $d$ ($a_J^{\ast}$)  & $\theta(^{\circ})$ &    $N_4$    &   $N_5$  & $R_{45}$     \\

\hline\hline
           
1 $m_J$   &       13       &        330       &     500     &    269    &   1.86   \\


100 $m_E$ &       13       &        330       &     500     &    445    &  1.12    \\


100 $m_E$ &      12.8     &        300       &     490     &    238    &  2.06    \\

50 $m_E$ &       12.8     &        295       &     495     &    297    &  1.67    \\

10 $m_E$ &       12.8     &        330       &     500     &    487    &  1.03   \\

\hline
\end{tabular}
\end{minipage}
\end{table}

Concerning the L4 Trojans, as seen in Table \ref{traveseMffp}, we notice in a couple of cases that $N_4$ is slightly smaller than the number of initial test Trojans (i.e. $=500$). This may be due to the fact that since the FFP is started with $d=12.8~a_J^{\ast}$ and having its perihelion just on Jupiter's orbit, it could pass by the L4 region from a distance, not up close, and consequently a tiny fraction of the local Trojans could be removed. Nevertheless, such a small decrease in the number of the L4 Trojans does not affect our results at all.

\subsubsection{FFP's inclination}

\begin{table}
\centering
\hspace{-2cm}
\begin{minipage}{7cm}
\caption{Similar to Table \ref{FFPinc}, but for the direct case of FFP traversing the L5 region. The results (left three columns) are shown for surviving test Trojans. The system's parameters are adopted to be $m_{FFP}=1~m_J$, $d=12.8~a_J^{\ast}$ and $\theta=330^{\circ}$.}      
\label{FFPinc2}
\begin{tabular}{c c c c}        

$i_{FFP}=0^{\circ}$~~~($z^p_{FFP}=0$)      \\
\hline\hline
$\Delta z^p_{FFP}$ ($a_J^{\ast}$)  &    $N_4$    &      $N_5$    &     $R_{45}$   \\
\hline

0                                  &   500       &       170     &     2.94    \\
\hline\hline\\

$i_{FFP}=10^{\circ}$~~~($z^p_{FFP}=-0.18a_J^{\ast}$)      \\
\hline\hline
$\Delta z^p_{FFP}$ ($a_J^{\ast}$)  &    $N_4$    &      $N_5$    &     $R_{45}$   \\
\hline

0                                  &  500       &    267         &    1.87   \\

\hline\hline\\

$i_{FFP}=20^{\circ}$~~~($z^p_{FFP}=-0.35a_J^{\ast}$)      \\
\hline\hline
$\Delta z^p_{FFP}$ ($a_J^{\ast}$)  &    $N_4$    &      $N_5$    &     $R_{45}$   \\ 

\hline
0                                  &   500      &      483     &  1.03     \\
\hline
0.1                               &   500       &      342     &  1.46     \\
\hline
0.2                               &   500      &       232      &  2.16      \\

\hline\hline\\

$i_{FFP}=30^{\circ}$~~~($z^p_{FFP}=-0.52a_J^{\ast}$)      \\
\hline\hline
$\Delta z^p_{FFP}$ ($a_J^{\ast}$)  &    $N_4$    &      $N_5$    &     $R_{45}$   \\ 
\hline

0.2                                &  500        &       387     &    1.29   \\
\hline
0.3                                &  500        &       305        &    1.64   \\

\hline\hline\\

$i_{FFP}=40^{\circ}$~~~($z^p_{FFP}=-0.66a_J^{\ast}$)      \\
\hline\hline
$\Delta z^p_{FFP}$ ($a_J^{\ast}$)  &    $N_4$    &      $N_5$    &     $R_{45}$   \\   
\hline
0.4                                &  500        &       345      &   1.45  \\
\hline
0.5                               &  500         &        233     &   2.15  \\

\hline\hline\\

$i_{FFP}=50^{\circ}$~~~($z^p_{FFP}=-0.79a_J^{\ast}$)      \\
\hline\hline
$\Delta z^p_{FFP}$ ($a_J^{\ast}$)  &    $N_4$    &      $N_5$    &     $R_{45}$   \\   
\hline
0.5                                &   500       &       345        &    1.45   \\
\hline
0.6                                &   500       &       264    &   1.89  \\

\hline\hline\\

$i_{FFP}=60^{\circ}$~~~($z^p_{FFP}=-0.87a_J^{\ast}$)      \\
\hline\hline
$\Delta z^p_{FFP}$ ($a_J^{\ast}$)  &    $N_4$    &      $N_5$    &     $R_{45}$   \\
\hline
0.6                               &   500       &         344        &    1.45   \\
\hline
0.7                               &   497     &         281    &    1.77  \\

\hline\hline\\

\end{tabular}
\end{minipage}
\end{table}

In analogy to the investigation of different inclinations $i_{FFP}$ of the FFP concerning the jumping Jupiter model in Sect. 2.4.2, here, we evaluate the capability of an inclined FFP to directly deplete the L5 Trojans. In order to achieve the largest perturbation of the FFP when it goes towards the L5 point, the three system parameters, $m_{FFP}$, $d$ and $\theta$, are respectively adopted to be $1~m_J$, $12.8~a_J^{\ast}$ and $330^{\circ}$ according to our simulations performed in the previous subsection. The results are presented in Table \ref{FFPinc2}.

We first consider the FFP starting with the initial conditions given in Eqs. (\ref{incliedX}) and (\ref{incliedV}). In this case, the FFP moves on the original trajectory without any elevation of its orbital plane along the $z$-axis, indicated by $\Delta z^p_{FFP}=0$. As shown in Table \ref{FFPinc2}, the leading-to-trailing number ratio of surviving Trojans decreases from $R_{45}=2.94$ at $i_{FFP}=0^{\circ}$ to $R_{45}=1.87$ at $i_{FFP}=10^{\circ}$. Furthermore, at $i_{FFP}=20^{\circ}$, the value of $R_{45}$ is nearly equal to 1, it means that the number asymmetry has almost disappeared. Therefore, the strength of the FFP's perturbation on the L5 Trojans weakens very quickly as $i_{FFP}$ becomes higher. We would like to remark that, for the same planar case of $i_{FFP}=0^{\circ}$, the number ratio $R_{45}=2.94$ given in Table \ref{FFPinc2} is much larger than the value of 1.86 from the control model. This is because the perturbation of the FFP is stronger due to a smaller $d=12.8~a_J^{\ast}$ adopted here.


Next, we lift up the FFP's orbital plane along the positive $z$-axis and then the perihelion of the FFP would be closer to the orbital plane of the Trojan population. This way, at high $i_{FFP}$, the FFP's perturbation on the L5 Trojans can be partly offset by increasing $\Delta z^p_{FFP}(>0)$. Table \ref{FFPinc2} shows that, given proper values of $\Delta z^p_{FFP}$, the unbiased number asymmetry of $R_{45}\sim1.6$ can always be achieved for a FFP with inclination $i_{FFP}$ up to as high as $60^{\circ}$. We then suppose that the chance of a FFP entering the Solar System at an inclination within such a wide range of $i_{FFP}=0$-$60^{\circ}$ should not be extremely low. As we only aim in providing a new mechanism to explain the number asymmetry problem of Jupiter Trojans, the possibility of a more inclined FFP with $i_{FFP}>60^{\circ}$ will not be explored any further.

Observations show that Jupiter Trojans have inclinations up to about $40^{\circ}$-$50^{\circ}$. A plausible explanation of the inclination distribution of the Trojans is the `chaotic capture'. During an early instability of the Solar System, the planetesimals were stirred up to high-inclination orbits by the giant planets, and captured into Jupiter's Lagrangian regions afterwards \citep{morb05,nesy13}. Here we wonder if a highly inclined FFP that once gravitationally influenced the Trojans could reproduce their large inclinations. So we carefully checked the orbits of surviving Trojans from our simulations, but no significant inclination excitation has been found, as all these objects end with inclinations $\lesssim 2^{\circ}$. We think this outcome is reasonable, otherwise the L4 and L5 swarms would have quite different inclination distributions because only the latter is strongly perturbed by the FFP in our model.

\section{Conclusions and discussion}  
In \cite{li2023} we investigated the number asymmetry of Jupiter Trojans. That dynamical model only took into account the celestial bodies from within our Solar System. Here, we extend this study by considering the invasion of a FFP from interstellar space. On one hand, the FFP could perturb Jupiter's motion and consequently affect the evolution of the Trojans asteroids in an indirect way. On the other hand, the FFP could directly interact with the Trojans themselves. For both cases, we explored the possibilities of explaining the unbiased L4-to-L5 number ratio of $R_{45}\sim1.6$.

In the indirect case, we consider a jumping-Jupiter that is accelerated via the close encounter with a FFP. We construct the invasion of a Jupiter-mass FFP on a parabolic orbit with perihelion outside Jupiter's orbit, and then investigate the FFP's gravitational effect on the outward migration of Jupiter. The results show that Jupiter's migration amplitude $\Delta a_J$ could vary in the range of (0, 0.3] AU depending on how deep the close encounter is; while the migration timescale $\Delta t$ is always of the order of 10 yr, determined by the fast pass-by process of the FFP. Accordingly, Jupiter's migration speed $\dot{a}_J=\Delta a_J/\Delta t$ could reach a value up to 0.03 AU/yr. Because $\Delta t$ is nearly a constant, the speed $\dot{a}_J$ can be solely characterised by $\Delta a_J$. Then, in the jumping-Jupiter model, we observe the evolution of the L4 and L5 test Trojans starting with symmetric resonant angles of $60^{\circ}+\Delta \sigma_0$ and $-60^{\circ}-\Delta \sigma_0$ respectively, where $\Delta\sigma_0$ is taken to be the same standard setting of $30^{\circ}$-$80^{\circ}$ as in \citet{li2023}. Given an equal number of initial test Trojans for each swarm, we find that, when $\Delta a_J=0.12$ AU, the resulting number ratio $R_{45}$ can be about 1.6. Furthermore, this ratio can continue to go up to a much larger value as $\Delta a_J$ increases. Then we are allowed to consider other $\Delta\sigma_0$ smaller than the standard setting, which means that the test Trojans are initially closer to the Lagrangian points and the final $R_{45}$ would be lower due to the more stable L5 swarm. Because in this situation, we can accordingly increase the migration speed $\dot{a}_J$ (i.e. by increasing $\Delta a_J$) to enhance the number asymmetry of the Trojans. When we finally reduce $\Delta\sigma_0$ to be as small as $0^{\circ}$-$50^{\circ}$, given $\Delta a_J=0.2$ AU, a number ratio of $R_{45}\sim1.6$ can also be obtained to explain the unbiased number asymmetry of Jupiter Trojans.


It is very important to point out the fundamental difference between our previous work and this one, concerning the jumping-Jupiter model. In \citet{li2023}, Jupiter's migration speed $\dot{a}_J$ is always lower than a critical value of $\dot{a}_J^{crit}=3.8\times10^{-3}$ AU/yr. Under that condition, although the L4 region contracts and the L5 region expands, both of them can exist. As a result, the faster migration of Jupiter would induce more stable L4 Trojans and fewer stable L5 Trojans, and accordingly the L4-to-L5 number ratio $R_{45}$ increases. But, in this work, the jumping-Jupiter has a speed of $>0.01$ AU/yr, which is higher than the value of $\dot{a}_J^{crit}$. As a consequence, during the outward migration of Jupiter, the two Lagrangian points L3 and L4 merge, and the L4 Trojans temporarily switch to the horseshoe orbits and move towards L3. When Jupiter's migration ceases within about 10 yr, the L4 point reappears and the surrounding objects come back to tadpole orbits but possess larger resonant amplitudes. So much differently from \citet{li2023}, the faster migration of Jupiter would induce fewer stable L4 Trojans due to the resonant amplitude enlargement, as a larger population of this swarm would be ejected eventually. As for the L5 Trojans, their tadpole orbits can always exist but would become considerably wider due to such fast migration of Jupiter. Taken together, although both the Trojan swarms have their resonant amplitudes increased, the L4 one suffers less excitation in the resonant amplitude and has more stable Trojans that could survive. Therefore, the resulting number ratio $R_{45}$ can be at the level of 1.6 or even larger. More interestingly, we find that $R_{45}$ can become infinite when the migration of Jupiter is fast enough, since in this case some L4 Trojans remain stable but all the L5 Trojans have escaped.

At this point, we have achieved a complete understanding of the effect of Jupiter's fast outward migration on the number asymmetry of the L4 and L5 Jupiter Trojan swarms, for both the cases of migration speeds $\dot{a}_J$ \textit{below} \citep{li2023} and \textit{above} (this work) the critical value of $\dot{a}_J^{crit}=3.8\times10^{-3}$ AU/yr. 
It must be mentioned that, the faster migration of Jupiter with $\dot{a}_J>\dot{a}_J^{crit}$ does not necessarily require the invasion of a FFP. This process could be attributed to any strong enough acceleration exerted on Jupiter; for instance, an extremely deep close encounter with some ice giant. 


However, the mechanism of jumping-Jupiter and L4/L5 region distortion always encounters the problem that, in the cases where we can obtain $R_{45}\sim1.6$, the simulated Trojans generally have quite larger resonant amplitudes than those of the real Trojans. As we discussed in \citet{li2023}, this issue may be tackled with some dissipation of energy, for example, the mutual collisions. To deal with the inconsistency of the resonant amplitude distribution, we further propose a second mechanism summarized as follows.

In the direct case, we assume that a FFP once traversed the L5 region and thus it could deplete the local Trojans, while the L4 Trojans at large distances were not affected. We find that due to the perturbation of a Jupiter-mass FFP, only 53.8\% of the L5 Trojans could survive after 1 Myr evolution, but all the L4 Trojans remain on tadpole orbits. This leads to a number ratio of $R_{45}=1.86$, which is potentially capable of explaining the observed asymmetry of $R_{45}\sim1.6$ as long as the FFP is a bit less massive or has a larger encounter distance to Jupiter. In addition, both the L4 and L5 swarms have considerable fractions of simulated Trojans with resonant amplitudes below $30^{\circ}$, similar to the real Jupiter Trojans. As a matter of fact, a similar mechanism was proposed by \citet{nesy13}. But in their model, the value of $R_{45}$ is found to be about 1.3, which could be due to a smaller invader of the L5 region, that is, an ice giant. 

Surely, in either the indirect case or the direct case, the effect of the FFP's invasion on the number ratio $R_{45}$ of Jupiter Trojans depends on its parameters, especially the mass $m_{FFP}$ and inclination $i_{FFP}$.
It is very interesting to find that in both cases, in order for the L4/L5 number asymmetry to be explained at the level of $R_{45}\sim1.6$, the FFP could have similar parameter ranges of $m_{FFP}$ from tens of Earth masses to a few Jupiter masses and $i_{FFP}$ as large as $40^{\circ}$. Given such wide ranges of $m_{FFP}$ and $i_{FFP}$, we suppose that the chance of a proper FFP entering our Solar System should not be extremely low.


\begin{acknowledgements}
    
This work was supported by the National Natural Science Foundation of China (Nos. 11973027, 11933001, 12150009), and National Key R\&D Program of China (2019YFA0706601). And part of this work was also supported by a Grant-in-Aid for Scientific Research (20H04617). We would also like to express our sincere thanks to the anonymous referee for the valuable comments.
      
\end{acknowledgements}

%

\begin{thebibliography}{}





\bibitem[Di Sisto et al. (2014)]{disi14}
Di Sisto, R. P., Ramos X. S., \& Beaug\'e, C. 2014, Icarus, 243, 287

\bibitem[Di Sisto et al. (2019)]{disi19}
Di Sisto, R. P., Ramos X. S., \& Gallardo, T. 2019, Icarus, 319, 828






\bibitem[Freistetter (2006)]{frei06} 
Freistetter, F. 2006, A\&A, 453, 353

\bibitem[Goulinski~\&~Ribak (2018)]{goul18} 
Goulinski, N., \& Ribak, E. N. 2018, MNRAS, 473, 1589

\bibitem[Grav et al. (2011)]{grav11} 
Grav, T., Mainzer, A. K., Bauer, J., et al. 2011, ApJ, 742, 40

\bibitem[Grav et al. (2012)]{grav12} 
Grav, T., Mainzer, A. K., Bauer, J. M., et al. 2012, ApJ, 759, 49


\bibitem[Holt et al. (2020)]{holt20}
Holt, T. R., Nesvorn\'y, D., Horner, J., et al. 2020, MNRAS, 495, 4085


\bibitem[Jewitt et al. (2004)]{jewi04}
Jewitt, D. C., Sheppard, S., \& Porco, C. 2004, in Jupiter: The Planet, Satellites and Magnetosphere, ed. F. Bagenal, T. Dowling, \& W. McKinnon (Cambridge: Cambridge Univ. Press), 263



\bibitem[Levison~\&~Duncan (1994)]{levi94} 
Levison, H. F., \& Duncan, M. J. 1994, Icarus, 108, 18

\bibitem[Li~\&~Adams (2016)]{lig16} 
Li, G., \& Adams, F. C. 2016, ApJL, 823, L3


\bibitem[Li et al. (2019)]{li2019} 
Li, J., Xia, Z. J., \& Zhou L., 2019, A\&A, 630, A68

\bibitem[Li~\&~Sun (2018)]{li2018} 
Li, J., \& Sun, Y.-S. 2018, A\&A, 616, A70

\bibitem[Li et al. (2023)]{li2023} 
Li, J., Xia, Z. J., Yoshida, F., et al. 2023, A\&A, 669, A68

\bibitem[Malhotra (1995)]{malh95}
Malhotra R., 1995, AJ, 110, 420



\bibitem[Marzari~\&~Scholl (1998)]{marz98} 
Marzari, F., \& Scholl, H. 1998, A\&A, 339, 278

\bibitem[Marzari~\&~Scholl (2000)]{marz00} 
Marzari, F., \& Scholl, H. 2000, Icarus, 232, 239





\bibitem[Melita et al. (2008)]{meli08}
Miret-Roig, N, Bouy, H, Raymond, S. N., et al. 2022, Nature Astronomy, 6, 89



\bibitem[Morbidelli et al. (2005)]{morb05}
Morbidelli, A., Levison, H. F., Tsiganis, K., \& Gomes, R. 2005, Nature, 435, 462


\bibitem[Murray-Clay~\&~Chiang (2005)]{murr05} 
Murray-Clay, R. A., \& Chiang, E. I. 2005, ApJ, 619, 623

\bibitem[Nakamura~\&~Yoshida (2008)]{naka08}
Nakamura, T., \&  Yoshida, F. 2008, Publ. Astron. Soc. Japan, 60, 293

\bibitem[Nesvorn\'y (2018)]{nesy08} 
Nesvorn\'y, D. 2018, ARA\&A, 56, 137



\bibitem[Nesvorn\'y et al. (2013)]{nesy13}
Nesvorn\'y, D., Vokrouhlick\'y, D., \& Morbidelli, A. 2013, ApJ, 768, 45

\bibitem[Nicholson (1961)]{nich61}
Nicholson, S. B. 1961, Astronomical Society of the Pacific Leaflets, 8, 239

\bibitem[Ogilvie~\&~Lubow (2006)]{ogil06} 
Ogilvie, G. I., \& Lubow, S. H. 2006, MNRAS, 370, 784


\bibitem[Parker et al. (2017)]{park17} 
Parker, R. J., Lichtenberg T., \& Quanz S. P. 2017, MNRAS, 472, L75





\bibitem[Shoemaker et al. (1989)]{shoe89} 
Shoemaker, E. M., Shoemaker, C. S., \& Wolfe, R. F. 1989, in Asteroids II, eds. R. P. Binzel, T. Gehrels, \& M. S. Matthews (Tucson: Univ. of Arizona Press), 487

\bibitem[Sicardy~\&~Dubois (2003)]{sica03}
Sicardy, B. \& Dubois, V. 2003, Celest. Mech. Dyn. Astron., 86, 321


\bibitem[Slyusarev (2013)]{slyu13}
Slyusarev I. G. 2013, Lunar Planet. Sci. 1719, 2223

\bibitem[Sumi et al. (2011)]{sumi11} 
Sumi, T. , Kamiya, K., Bennett, D. P., et al. 2011, Nature, 473, 349

\bibitem[Szab\'o et al. (2007)]{szab07} 
Szab\'o, Gy. M.,  Ivezi\'c, \v{Z}., Juri\'c, M., \& Lupton, R. 2007, MNRAS, 377, 1393



\bibitem[Varvoglis et al. (2012)]{varv12}
Varvoglis, H., Sgardeli, V., \& Tsiganis, K. 2012, Celest. Mech. Dyn. Astron., 113, 387





















































 



   \end{thebibliography}
%

\end{document}